\documentclass[11pt,preprint]{aastex}
\usepackage{epsf}

\newcommand{\etal}{{et al.\hskip 3pt}}

\newcommand{\kmsmpc}{\kms\;{\rm Mpc}^{-1}}
\newcommand{\lya}{Ly$\alpha$\ }
\newcommand{\kms}{\;{\rm km}\,{\rm s}^{-1}}

\newcommand{\hkpc}{h^{-1}{\rm kpc}}
\newcommand{\hmpc}{h^{-1}{\rm Mpc}}
\newcommand{\mpc}{{\rm Mpc}}
\newcommand{\lcdm}{$\Lambda$CDM}

\newcommand{\mbj}{{M_{b_{\rm J}}}}
\newcommand{\mbjs}{{M^*_{b_{\rm J}}}}
\newcommand{\vpair}{v_{12}(r)}
\newcommand{\spair}{\sigma_{12}(r)}
\newcommand{\xigm}{\xi_{\rm gm}}
\newcommand{\xigg}{\xi_{\rm gg}}
\newcommand{\ximm}{\xi_{\rm mm}}
\newcommand{\vx}{{\bf x}}
\newcommand{\vecr}{{\bf r}}
\newcommand{\bxi}{b_\xi}
\newcommand{\bxix}{b_{\xi\times}}
\newcommand{\rx}{r_{\times}}
\newcommand{\mstel}{M_{\rm stel}}
\newcommand{\stf}{S_{3{\rm f}}}

\slugcomment{Submitted to the Astrophysical Journal, December 16, 2002}

\shortauthors{Weinberg et al.}
\shorttitle{Galaxy Clustering and Bias}
\singlespace

\begin{document}

\title{Galaxy Clustering and Galaxy Bias in a $\Lambda$CDM Universe}

\author{David H. Weinberg}
\affil{Astronomy Department, Ohio State University, Columbus, OH 43210;
dhw@astronomy.ohio-state.edu}
\author{Romeel Dav\'e\altaffilmark{1} }
\affil{Steward Observatory, University of Arizona, Tucson, AZ 85721;
rad@as.arizona.edu}            
\author{Neal Katz}
\affil{Astronomy Department, University of Massachusetts, Amherst, MA 01003;
nsk@kaka.astro.umass.edu}
\and
\author{Lars Hernquist}
\affil{Harvard-Smithsonian Center for Astrophysics, Cambridge, MA, 02138;
lars@cfa.harvard.edu}

\altaffiltext{1}{Hubble Fellow}

\begin{abstract}
We investigate galaxy clustering and the correlations between galaxies and
mass in the $\Lambda$CDM cosmological model (inflationary cold dark matter
with $\Omega_m=0.4$, $\Omega_\Lambda=0.6$, $h=0.65$, $n=0.95$, $\sigma_8=0.8$),
using a large, smoothed particle hydrodynamics simulation (SPH, with 
$2\times 144^3$ particles in a $50\hmpc$ cube).  Simulated galaxies can be
unambiguously identified as clumps of stars and cold gas a few kpc to a few
tens of kpc across, residing in extended halos of hot gas and dark matter;
the space density of the resolved galaxy population at $z=0$ corresponds to 
that of observed galaxies with luminosity $L \ga L_*/4$.  We investigate the
galaxy correlation function, the pairwise velocity dispersion and mean pairwise
velocity, and the second and third moments of counts-in-cells; we also 
investigate the galaxy-mass correlation function and the average extended mass 
distributions around galaxies, both of which can be measured via galaxy-galaxy
lensing.  For the most part, the predicted biases between galaxies and dark
matter lead to good agreement with current observations, including: (1) a 
nearly
constant comoving correlation length from $z=3$ to $z=0$ for mass-selected 
galaxy samples of constant comoving space density; (2) an rms bias factor 
$b_\sigma \approx 1$ at $z=0$; (3) a scale-dependent bias on small scales 
that
transforms the curved dark matter correlation function into a nearly 
power-law galaxy
correlation function; (4) galaxy pairwise dispersion and hierarchical skewness
ratio $S_3$ in good agreement with observed values, and lower than values for
the dark matter by $\sim 20\%$; (5) a ratio of galaxy-galaxy to galaxy-mass 
correlation functions consistent with recent measurements from the Red Cluster
Sequence survey; and (6) a mean excess mass $\Delta M(260\hkpc)$ approximately
proportional to galaxy baryon mass $M_b$, in agreement with estimates from the
Sloan Digital Sky Survey (SDSS).  All of these clustering properties vary with
galaxy baryon mass and, more strongly, with the age of a galaxy's stellar
population.  The predicted dependences are in good qualitative agreement with
the observed dependence of galaxy clustering and the galaxy-mass correlation
function on galaxy type.  The predicted ratio $\Delta M(260\hkpc)/M_b$ is 
lower
than the SDSS estimates by a factor of $\sim 1.5-2$ for galaxies with 
$M_b \ga 2\times 10^{11}M_\odot$.  A test with a higher resolution (smaller
volume) simulation suggests that this discrepancy is mostly a numerical 
artifact; if so, then the SDSS weak lensing comparison leaves little room for
feedback or other astrophysical processes to reduce the stellar masses of 
luminous galaxies, at least given our adopted cosmological parameters.  On the
whole, our results show that the $\Lambda$CDM model and the galaxy formation 
physics incorporated in the SPH simulation give a good account of observed
galaxy clustering, but anticipated improvements in clustering and weak lensing
measurements will soon test this picture in much greater detail.
\end{abstract}


\section{Introduction}\label{sec: intro}

The clustering of galaxies has long been an essential testing ground
for cosmological models and the theory of galaxy formation, with comparisons
between predicted and observed clustering driving much of the progress in 
cosmology during the 1970s and 1980s.  Such comparisons provided early
support for gravitational instability as the central mechanism of
structure formation \citep[e.g.,][]{davis77},
for initial conditions with a power spectrum redder than
white noise \citep[e.g.,][]{gott75,gott77},
for a ``bottom up'' rather than ``top down'' sequence of
gravitational clustering \citep[e.g.,][]{white83,davis85,fry85},
for an approximately Gaussian distribution of primordial
fluctuations \citep[e.g.,][]{gott89,weinberg92,bouchet93},
and for a power spectrum significantly redder than that predicted
by the ``standard'' ($\Omega_m=1$, $h\equiv H_0/100\kmsmpc=0.5$)
cold dark matter scenario \citep[e.g.,][]{efstathiou90,maddox90,park94}.
The growth of galaxy redshift surveys has led to measurements of increasing
precision and detail, and cosmological N-body simulations have
developed into a powerful tool for calculating the gravitational
clustering of collisionless dark matter from specified initial
conditions.  The main obstacle to drawing stronger inferences from
the data is the dependence of theoretical predictions on {\it both}
the relatively
straightforward physics of gravitational clustering, which
largely determines the distribution of dark matter, and the more
complex physics of galaxy formation, which determines the relation 
between galaxies and mass, often referred to generically as ``bias.''

In this paper, we study galaxy clustering and galaxy bias in a large
($2\times 144^3$ particles, $50\hmpc$ cube), 
smoothed particle hydrodynamics (SPH) simulation of
a low density, 
inflationary cold dark matter universe with a cosmological constant
(\lcdm).  Our goals are, first, 
to see whether the currently leading cosmological
model and ``standard'' galaxy formation physics as incarnated in our SPH
simulation can reproduce existing observations, and, second, 
to give guidance for 
the physical interpretation of those observations and for the design
of more demanding tests of the theoretical predictions.
The N-body+hydro approach to {\it ab initio} predictions of galaxy bias
has a substantial history, 
beginning with dissipative ``sticky particle'' simulations by \cite{carlberg89}
and continuing in the early 1990s with simulations using
SPH \citep{katz92,evrard94} or Eulerian grid hydrodynamics \citep{cen92}
to model galaxy formation in a representative (but small) cosmological
volume.  
The present paper is a direct descendant of Katz et al.\ (1992),
but we study a different cosmological model,
we incorporate improvements to the physical treatment of radiative
cooling and star formation 
\citep[][hereafter KWH]{katz96}, and, above all, we take
advantage of the parallel implementation of TreeSPH \citep{dave97}
and advances in computer technology to simulate a volume more than 90
times larger than that of Katz et al.\ (1992), 
at similar resolution.  Our analysis overlaps significantly
with other recent studies using large volume SPH 
\citep{pearce99,pearce01,yoshikawa01} or Eulerian hydrodynamics \citep{cen00}
simulations.  Relative to these investigations, we have higher mass and/or
spatial resolution and a somewhat smaller simulation volume,
as discussed in \S\ref{sec: methods} below.

Two other approaches to {\it ab initio} predictions of bias have
gained currency in recent years: high-resolution, collisionless
N-body simulations that identify galaxies with ``subhalos'' in the
dark matter distribution \citep[e.g.,][]{colin99,kravtsov99},
and a hybrid method that combines N-body
simulations of the dark matter component with semi-analytic treatments
of the galaxy formation physics 
\citep[e.g.,][]{kauffmann97,governato98,kauffmann99a,kauffmann99b,benson00a,
benson00b,benson01,hatton02}.  
We will discuss comparisons of our results to those
from other hydrodynamic simulations and from the high-resolution N-body
and hybrid approaches as they arise.  We carry out a detailed comparison
between our simulation and the semi-analytic model of
\cite{benson00a} in a separate paper that focuses on the
``halo occupation distribution'' (HOD) predicted by the two methods
\citep{berlind02b}.
The HOD description can be used to calculate many different
galaxy clustering statistics, and it helps explain the origin of
bias in a physically intuitive manner (see \citealt{berlind02} and
references therein).
Based on this comparison, we conclude that our SPH approach and
Benson et al.'s (\citeyear{benson00a}) semi-analytic method should
yield similar predictions for most galaxy clustering statistics.

Here we focus mainly on the ``classic'' measurements of galaxy clustering ---
the two-point correlation function, variance of galaxy counts, 
and pairwise velocity moments --- and on one of the simplest measures
of higher-order clustering, the third moment of counts-in-cells.
Large redshift surveys like the 2dF Galaxy Redshift Survey (2dFGRS)
and the Sloan Digital Sky Survey (SDSS) now allow precise measurements
of these quantities for multiple classes of galaxies, defined by
luminosity, color, morphology, or spectral type (see, e.g.,
\citealt{norberg01,norberg02b,zehavi02}, and references therein).
While we cannot model these variations with galaxy type in detail ---
our simulation volume is far smaller than these surveys, and we do not resolve
the morphology of the simulated galaxies --- we can examine the
predicted trends of clustering with baryon mass, stellar population
age, and local environment.  We concentrate on present-day clustering,
but we also compute the evolution of the two-point function, which can
be compared to results from deep redshift surveys and to studies
of Lyman-break galaxies at $z\approx 3$ 
\citep[e.g.,][]{adelberger98,adelberger02}.
The discovery that the clustering of Lyman-break galaxies is similar
to that of $L_*$ galaxies at $z=0$, despite the weaker expected
clustering of the underlying mass distribution, provides strong evidence
that the bright galaxy population was highly biased at $z\approx 3$,
even if galaxies roughly trace mass today.

Galaxy-galaxy weak lensing is an important new observational probe
of the relation between galaxies and dark matter, measuring the
galaxy-mass cross-correlation function and the extended mass distributions
around galaxies of different types and in different environments.
We will devote considerable attention to modeling recent observations
of this phenomenon.  The closest similar efforts are those of
\cite{white01}, who present predictions from a hydrodynamic simulation
at $z=1$ and $z=0.5$, and \cite{guzik01} and \cite{yang02}, who analyze the
N-body+semi-analytic galaxy distribution of \cite{kauffmann99a}.

One strength of the hydrodynamic simulation approach is that it predicts
properties of the intergalactic medium (IGM) in addition to the
galaxy population.  We have used the simulation analyzed here to model
the phase distribution of baryons in the present day universe
\citep{dave01}, the X-ray background \citep{croft01},
X-ray emission from galaxy groups \citep{dave02},
and X-ray absorption by the diffuse IGM \citep{chen02}.
In combination with other simulations, we have used it to study 
cooling radiation and sub-millimeter emission from young galaxies
(\citealt{fardal01}, \citeyear{fardal02}),
the relative importance of mergers and smooth accretion in
galaxy assembly \citep{murali02}, and
the correlations between galaxies and the \lya\ forest at high redshift
\citep{kollmeier02}.  The observed distributions and correlations
of galaxy properties, in particular the luminosity function and
the Tully-Fisher (\citeyear{tully77}) relation, are also essential
tests of the galaxy formation model, and any discrepancies with these
observations can provide insight into the model's failings.
We will consider these characteristics of the galaxy population
in a separate paper (Katz et al., in preparation), though the
comparison to observed galaxy-galaxy lensing measurements will
bring up some of the same issues here.  We proceed to a description
of our numerical methods in \S\ref{sec: methods}, to predictions
of galaxy clustering in \S\ref{sec: clust}, and to the galaxy-mass
correlation function and comparison to lensing measurements in
\S\ref{sec: gdm}. We recap our findings and discuss their implications
in \S\ref{sec: disc}.

\section{Simulating Galaxy Formation}\label{sec: methods}

\subsection{Simulation}\label{sec: sim}

We employ a hydrodynamic simulation of a \lcdm\ model with
$\Omega_m=0.4$, $\Omega_\Lambda=0.6$, $h=0.65$, $\Omega_b=0.02h^{-2}=0.0473$,
inflationary spectral index $n=0.95$, 
and a power spectrum normalization $\sigma_8=0.80$.
This model is consistent with
most available observational constraints \citep{bahcall99,jaffe00}, including
COBE normalization \citep{bennett96},
cluster masses \citep{eke96}, 
cosmic shear (\citealt{hoekstra02}, and references therein),
the deuterium abundance 
in high-redshift Lyman limit systems 
\citep{tytler96,burles97,burles98}, the Hubble 
diagram of Type Ia supernovae
\citep{riess98,perlmutter99}, and the flux power spectrum of the 
\lya forest \citep{croft99,croft02,mcdonald00}.

Our initial conditions represent a random $50\hmpc$ (comoving) cube, 
realized with 
$144^3$ gas and $144^3$ dark matter particles in a triply periodic volume,
yielding particle masses
of $m_{\rm gas}=8.5\times 10^8 M_\odot$ and $m_{\rm dark}=6.3\times 10^9
M_\odot$, respectively.  
These initial conditions are evolved
from redshift $z=49\rightarrow 0$ using Parallel
TreeSPH \citep{dave97}.  
We soften the gravitational force via a $10\hkpc$
(comoving) spline kernel, equivalent to a Plummer softening of $7\hkpc$.
For particle timesteps we use the criteria described by KWH
and \cite{quinn97}, setting the tolerance parameter $\eta$ (defined
in these papers) to 0.4.
The simulation was evolved to $z=0.5$ on the
Cray T3E at SDSC and finished on the SGI Origin 2000 at NCSA.

We include radiative cooling for primordial composition gas as
described by KWH.  The tests in \cite{weinberg97} show that photoheating
by an ionizing background artificially suppresses the formation 
of low mass galaxies in a simulation of this resolution, but that a
simulation evolved without an ionizing background produces much the same 
galaxy
population (above its resolution limit) as a higher resolution simulation
with an ionizing background.  Though \cite{weinberg97} considered 
an $\Omega_m=1$ CDM model rather than \lcdm, we expect their conclusions
to carry over to this model, and we therefore evolve the simulation
with no ionizing background.

We incorporate star formation using the algorithm described and
tested by KWH.  This algorithm converts cold, dense gas into 
collisionless stars at a rate governed by the local dynamical and
cooling timescales, returning supernova feedback as thermal energy
to the surrounding medium.  The tests in KWH and \cite{weinberg97} show
that the resulting galaxy population is insensitive to the one free
parameter of the algorithm and is similar to that identified using cold
gas in a simulation without star formation; it is gravity, gas
dynamics, and radiative cooling that determine where galaxies form and
how massive they will be.  A radical change to the star formation 
or feedback algorithm might alter the galaxy population in a significant
way, but provided that it gave the same {\it relative} masses and
ages of the galaxies (more precisely, that it preserved their rank order),
it would not affect the clustering results presented here, except
for the absolute values of mass-to-light ratios discussed in \S\ref{sec: gdm}.

Our numerical approach is similar to those used by 
\citeauthor{pearce99} (\citeyear{pearce99}, \citeyear{pearce01})
and \cite{yoshikawa01}, but we have opted for a somewhat smaller
volume and larger particle number, and consequently higher resolution.
Pearce et al.\ use $128^3$ particles in a $70\hmpc$ cube, so our
mass resolution is higher by $(70/50)^3 (144/128)^3 = 3.9$
(though the ratio of SPH particle masses is only 2.9 because
Pearce et al.\ use a lower value of $\Omega_b$).
The gravitational softening of the Pearce et al.\ simulation is
similar to ours at $z=0$ ($10\hkpc$ vs. our $7\hkpc$), but 
larger at higher redshift, since they keep the softening fixed
in physical coordinates at $z<2.5$.  The mass resolution of the
\cite{yoshikawa01} simulation is similar to that of Pearce et al.\
($128^3$ particles in a $75\hmpc$ cube), and the (Plummer equivalent)
gravitational softening is $41\hkpc$, constant in comoving
coordinates.  The most important methodological difference is that
we convert cold gas into collisionless stars, while Pearce et al.\
and Yoshikawa et al.\ ``decouple'' cold gas from hot gas when computing
SPH densities but leave the cold gas subject to SPH forces and
dissipation.  Despite the differences in numerical resolution, input
physics, and implementation details, we find quite good agreement
with these previous investigations where our results overlap,
as discussed in \S\ref{sec: clust}. 
The cosmological parameters of the three $\Lambda$CDM simulations are similar
but not identical. (Pearce et al.\ investigate an $\Omega_m=1$ model
in addition to $\Lambda$CDM.)

\cite{cen00} have investigated galaxy clustering and bias in the
$\Lambda$CDM model with a $512^3$ Eulerian mesh simulation of a 
$100\hmpc$ volume.  The initial mass resolution of this simulation is higher
than ours, but the force resolution is much lower, with individual
mesh cells $\approx 200\hkpc$.  Because the simulation does not
track individual galaxies within high density regions, 
Cen \& Ostriker examine stellar mass weighted clustering
statistics rather than the galaxy number weighted statistics that
we investigate here.  Nonetheless, our conclusions about galaxy
bias on the scales $r\ga 1\hmpc$ resolved by the Cen \& Ostriker
simulation are qualitatively similar, as discussed in \S\ref{sec: clust}. 

\begin{figure}
\centerline{
\epsfxsize=6.0truein
\epsfbox[20 20 575 575]{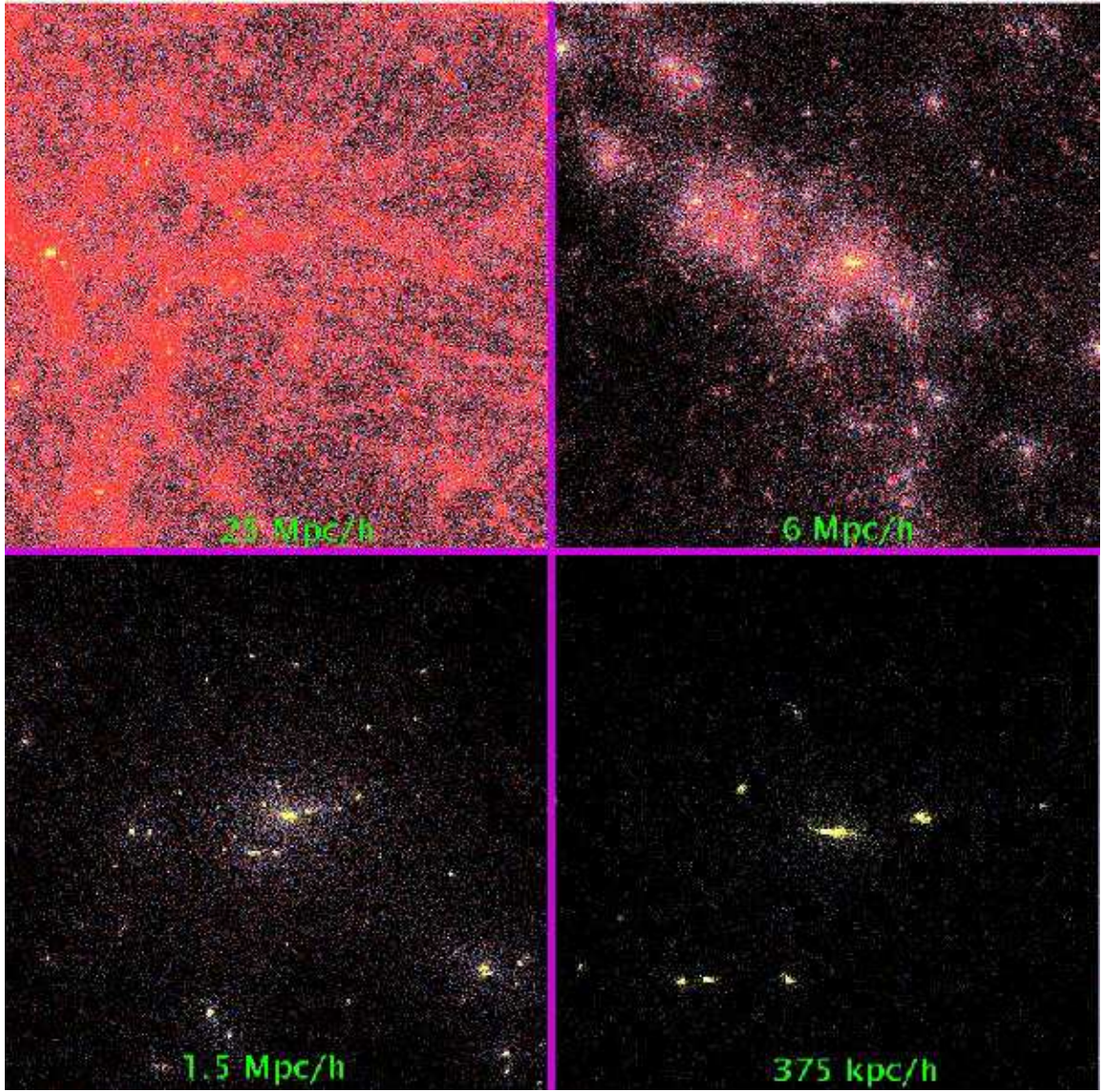}
}
\caption {Projections through the 
simulation volume at $z=0$.  Blue particles are
dark matter, red are gas, and yellow are baryonic particles where star
formation has occurred.  The size of each view is indicated.  The expanded
views are centered on the largest galaxy, seen near the left edge of 
the upper left panel. 
All panels show the projection through the entire simulation volume.}
\label{fig:tipsy}
\end{figure}

\subsection{Galaxy Identification}\label{sec:gal}

Figure~\ref{fig:tipsy} illustrates the state of the simulation at
$z=0$, showing dark matter, gas, and star particles projected through the
entire volume in a quarter of the box,
and in expanded views 6, 1.5, and 0.375$\hmpc$ on a side centered on 
the most massive system in our volume.  The star 
particles are clumped into groups of extremely high overdensity, from
a few kpc to a few tens of kpc across.  Dark halos with the mass scale
of galaxy groups and clusters contain many such dense clumps in addition
to a smooth distribution of hot gas.  Figure~\ref{fig:tipsy} shows
that there is virtually no ambiguity in identifying the galaxies in this
simulation; one simply needs an automated algorithm that picks out
the distinct clumps of stars and cold gas.  The algorithm that we use is
Spline Kernel Interpolative DENMAX
\footnote{\tt http://www-hpcc.astro.washington.edu/tools/SKID/}
(SKID), which selects gravitationally bound groups of particles that 
are associated with a common local density maximum (see KWH).
SKID selects essentially the same population of objects that one
would pick out by eye from a representation like Figure~\ref{fig:tipsy}.

\begin{figure}
\centerline{
\epsfxsize=3.5truein
\epsfbox[105 425 460 720]{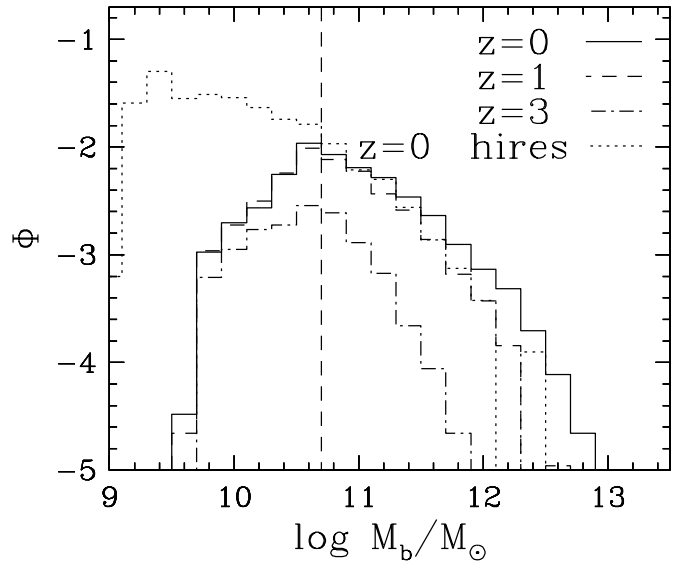}
}
\caption {The baryonic mass function of SKID-identified galaxies at redshifts
$z=0,1,3$ (solid, dashed, and dot-dashed histograms, respectively)
in the $(50\hmpc)^3$ simulation.
The dotted histogram shows the $z=0$ mass function from a higher
resolution simulation of a smaller volume.  Solid and dotted histograms
diverge below $M_b=5\times 10^{10}M_\odot$ (vertical dashed line),
corresponding to $60m_{\rm gas}$ in the large volume simulation,
which we adopt as our threshold for identifying resolved galaxies.
}
\label{fig:massfun}
\end{figure}

Figure~\ref{fig:massfun} shows the baryonic mass function of the simulated
galaxy population at $z=3$, 1, and 0, where $M_b$ represents
the total mass of a galaxy's stars and associated cold ($T<30,000\;$K) gas.
The number of galaxies steadily increases
with time, though the rate of new galaxies forming is dramatically
slower for $z=1\rightarrow 0$ than for $z=3\rightarrow 1$.
We will discuss the
$z=0$ luminosity function and Tully-Fisher relation
in a separate paper (Katz et al., in preparation),
and we have elsewhere
compared predictions of the $z=3$ luminosity function from
higher resolution simulations (of smaller volumes) to measurements
of the Lyman-break galaxy population (\citealt{weinberg99}, 
\citeyear{weinberg02}).
The dashed vertical line in Figure~\ref{fig:massfun} marks the mass
of 60 SPH particles, and all of the simulated mass functions turn over
shortly below this threshold.  Our tests using a suite of simulations
of different resolution show that galaxies below this threshold have
underestimated masses or may be missed entirely, while the masses
and locations of galaxies above this threshold are generally 
robust (see \citealt{weinberg99}).  
The dotted line shows the mass function at $z=0$ from a
higher resolution ($\epsilon=3\hkpc$, $m_{\rm gas}=1.04\times 10^8M_\odot$), 
smaller volume ($22.22\hmpc$ cube)
simulation having $2\times 128^3$ particles and the same cosmological 
parameters, analyzed in the same way; a comparison with the solid
line clearly shows that the turnover 
below the demarcated mass limit is due to numerical effects.
We therefore adopt a baryon mass of
$M_b =  60 m_{\rm gas} = 5\times 10^{10} M_\odot$
as our completeness limit, and we analyze the clustering only of
galaxies above this threshold.  Note that the mass function 
comparison also suggests
that the lower resolution simulation may overestimate the baryon
masses of the largest galaxies, a point that we will return to in 
\S\ref{sec: gdm}.

Out of 4632 total SKID-identified galaxies at $z=0$, 2571 satisfy our
$M_b\geq 60 m_{\rm gas}$
completeness criterion.  At $z=1, 2, 3$ and 4, the corresponding
fractions are 2011/4012,
1034/2488, 420/1177, and 104/384.  The sample of all 
galaxies with $M_b \geq 60 m_{\rm gas}$
will be referred to as our ``complete sample" of galaxies.
At $z=0$, the space density of the complete sample, 
$2571/(50\hmpc)^3 = 0.02 h^3 {\rm Mpc}^{-3}$, corresponds to that
of observed galaxies brighter than about $L_*/4$ \citep{blanton01},
where $L_*$ is the characteristic luminosity of the
\cite{schechter76} function fit.

\section{Galaxy Clustering}\label{sec: clust}

\begin{figure}
\centerline{
\epsfxsize=6.5truein
\epsfbox[10 105 545 685]{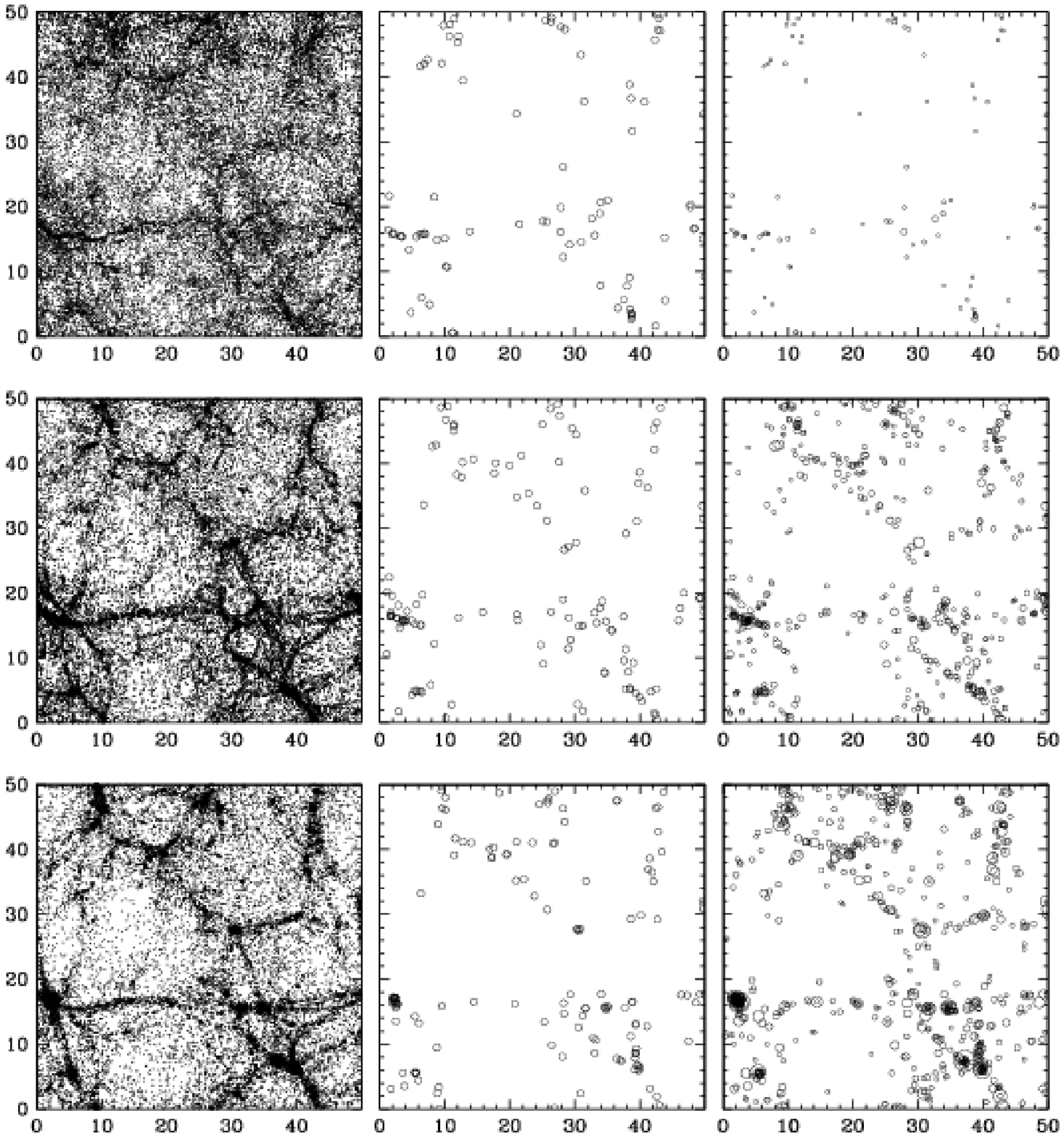}
}
\caption {
Structure in a $10\hmpc$ thick slice of the $50\hmpc$ cube at $z=3$
(top), $z=1$ (middle), and $z=0$ (bottom).  The left column shows
a randomly selected 5\% of the dark matter particles.  The middle column
shows the 420 resolved galaxies at $z=3$ and the 500 most massive
galaxies at $z=1$ and $z=0$.  The right column shows all resolved
galaxies at the three redshifts, with the area of each point proportional
to the galaxy baryon mass and the smallest points corresponding to
our resolution threshold of $5\times 10^{10}M_\odot$.
}
\label{fig:evol}
\end{figure}

Figure~\ref{fig:evol} illustrates the evolution of structure in a $10\hmpc$
thick slice of the simulation volume, running from $z=3$ (top) to 
$z=1$ (middle) to $z=0$ (bottom).  The left hand panels show the dark 
matter distribution, which exhibits the usual behavior seen in
N-body simulations: a steady increase of contrast between growing
filaments and emptying voids, and fragmentation of filaments into
increasingly massive clumps.  The middle column shows the clustering
of a subset of galaxies with nearly constant comoving space density,
namely the 420 resolved galaxies at $z=3$ and the 500 most massive galaxies
at $z=1$ and $z=0$.  We will refer to this subset below as our
``$L_*$ sample.''  The evolution of galaxy clustering is markedly
different from that of the dark matter.  The network of filaments
and voids is already present at $z=3$, with nearly the full contrast
that it achieves by $z=0$.  The principal change in the clustering
of this sample is on small scales, with the collapse of some extended
structures into tight clumps.  

The right column of Figure~\ref{fig:evol} shows the full resolved galaxy
population at the three redshifts, with a point size that reflects the 
galaxy baryon mass.  Here one can see the increase in the number of
resolved galaxies (from 420 to 2011 to 2571), the growth in the 
average mass of these galaxies, and the preferential location of the
most massive galaxies in dense systems at $z=0$.  The changing number
density and size of points makes it difficult to gauge the evolution
of clustering strength visually, but we will see below that the
correlation function of this complete galaxy sample stays nearly constant
from $z=3$ to $z=0$.

\subsection{The Two-Point Correlation Function}
\label{sec:corrall}

\begin{deluxetable}{lccl}
\footnotesize
\tablecaption{Real-space correlation function parameters $r_0$ and $\gamma$ 
from our simulation and recent redshift surveys.\label{tab:r0gamma}}
\tablewidth{0pt}
\tablehead{
\colhead{Survey} & 
\colhead{$r_0$} & 
\colhead{$\gamma$} &
\colhead{Reference}
} 
\startdata
All Simulated & $3.96\pm 0.30$ & $1.78\pm 0.05$ & \\
$L_\star$ Simulated & $4.52\pm 0.38$ & $2.00\pm 0.08$ & \\
200 Most Massive & $5.20 \pm 0.52$ & $2.05 \pm 0.09$ & \\
\hline \\
Stromlo-APM & $5.1\pm 0.2$ & $1.71\pm 0.05$ & \citet{loveday95} \\
CfA2/SSRS2\tablenotemark{a} & 5.8 & 1.8 & \cite{marzke95} \\
LCRS & $5.06\pm 0.12$ & $1.86\pm 0.034$ & \citet{jing98} \\
Durham/UKST & $5.1\pm 0.3$ & $1.6\pm 0.1$ & \citet{ratcliffe98} \\
ESP & $4.15\pm 0.2$ & $1.67\pm 0.08$ & \citet{guzzo00} \\
2dFGRS\tablenotemark{b} & $4.9\pm 0.3$ & $1.71\pm 0.06$ & \citet{norberg01} \\
SDSS\tablenotemark{c} & $6.1\pm 0.2$ & $1.75\pm 0.03$ & \citet{zehavi02} \\
PSCz\tablenotemark{d} & $3.7$ & $1.69$ & \citet{jing02} \\
\enddata
\tablenotetext{a}{Based on figure 3 of the \cite{marzke95} analysis of
CfA2 and SSRS2.}
\tablenotetext{b}{We quote values for the volume-limited sub-sample
with $-20.5 < \mbj < -19.5$, close to the value of $L_*$ 
found by \cite{norberg02a} for the 2dFGRS.}
\tablenotetext{c}{We quote values for the full flux-limited sample.
For a volume limited sample with $-21.5 < M_{r} < -20.0$, close to
the \cite{blanton01} value of $L_*$ for the SDSS, \cite{zehavi02} find
$r_0=6.3 \pm 0.8$ and $\gamma = 1.80 \pm 0.09$.}
\tablenotetext{d}{Note that these are IRAS-selected galaxies and
therefore preferentially late types.}
\end{deluxetable}

The two-point correlation function is the most well studied measure
of galaxy clustering, in part because it is relatively straightforward
to estimate from angular or redshift survey data, in part because it
contains complete statistical information on fluctuations in the linear
regime if these fluctuations are Gaussian, and, perhaps most of all,
because its observed form is remarkably simple.  For most galaxy samples,
the correlation function is well described by a power law,
\begin{equation}\label{eqn: r0gamma} 
\xi(r) = \Bigl( \frac{r}{r_0} \Bigr)^{-\gamma},
\end{equation} 
on scales $r \sim 0.01 - 10\hmpc$ \citep{totsuji69,peebles74,gott79}.
However, the values of $r_0$ and, to a lesser extent, $\gamma$,
depend on the luminosity, color, and type of the sample galaxies
(e.g., \citealt{guzzo97,norberg01,norberg02b,zehavi02}, 
and numerous references therein).
We list determinations of $r_0$ and $\gamma$ from a number of recent
redshift surveys in Table~\ref{tab:r0gamma}.
While the varying values of $r_0$ and $\gamma$ partly reflect
statistical uncertainties, they arise mainly from the different 
selection criteria of the samples from which they are measured.
In most cases, the values quoted in Table~\ref{tab:r0gamma} are
derived from flux-limited galaxy samples, and the weighting of galaxies
of different luminosities depends on the specific procedure adopted
for estimating $\xi(r)$.  The quoted values for the 2dFGRS are
derived from a volume-limited sample of galaxies with absolute
magnitudes close to the characteristic magnitude $\mbjs$
obtained for 2dFGRS galaxies.  The quoted values for the SDSS 
are from a flux-limited sample, but they are similar to those
obtained for a volume-limited sample with absolute magnitudes
close to $M^*_r$.  The difference between the 2dFGRS and SDSS
correlation lengths, $4.9 \pm 0.3\hmpc$ vs.\ $6.1 \pm 0.2\hmpc$
(or $6.3 \pm 0.8\hmpc$ for the volume-limited sample) presumably
reflects the difference between blue and red selection.

Early N-body simulations showed that gravitational clustering 
produces an approximate power-law $\xi(r)$, especially if the 
initial power spectrum has a slope $n\approx -1$ \citep{gott79b}.
However, the much higher precision correlation functions calculated
with modern simulations show significant departures from a power-law
$\xi(r)$ for CDM or power-law initial spectra 
(e.g., \citealt{jenkins98} and references therein).
The origin of these departures can be understood in analytic 
models of matter clustering \citep{hamilton91,peacock96,ma00,seljak00,
scoccimarro01}.
Thus, the observed power-law form of $\xi(r)$ requires either
special features in the primordial power spectrum or scale-dependent
bias between the galaxy and dark matter correlation functions
at the present day.

\begin{figure}
\centerline{
\epsfysize=6.0truein
\epsfbox[125 75 460 720]{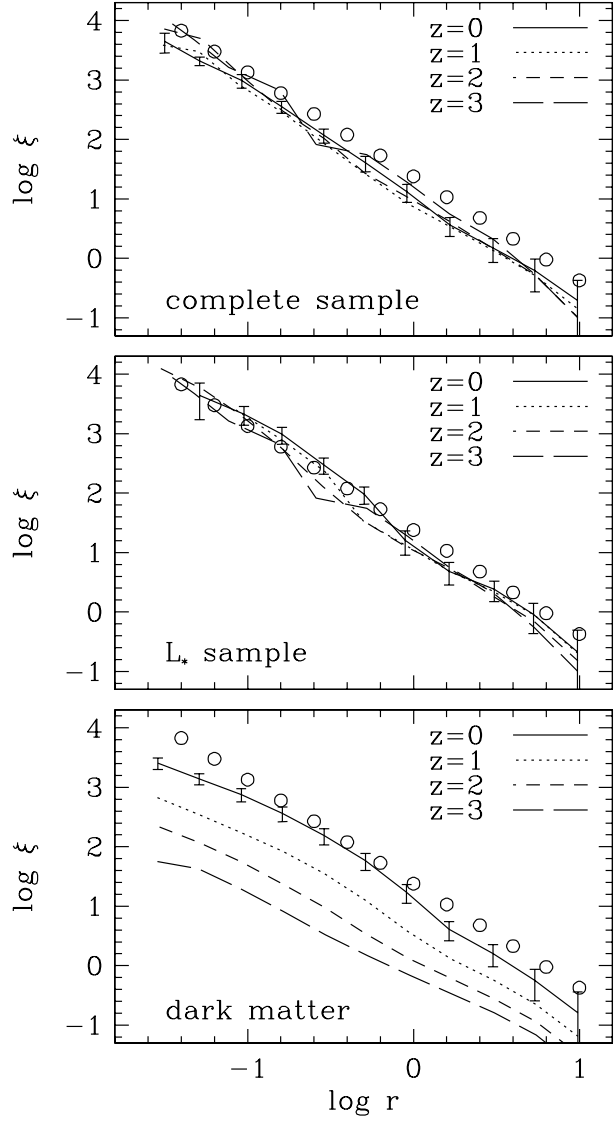}
}
\caption { Correlation functions $\xi(r)$ for the complete galaxy sample (top
panel), the $L_\star$ galaxy sample (middle), and 
the dark matter (bottom), at redshifts $z=0,1,2,3$.  
Jackknife error bars are plotted for the $z=0$ correlation function.
Open circles show a power law with the parameters measured from the
SDSS by \cite{zehavi02}, $r_0=6.1\hmpc$ and $\gamma=1.75$.
}
\label{fig:corrall}
\end{figure}

Figure~\ref{fig:corrall} illustrates the evolution of $\xi(r)$ in our
simulation from
$z=3$ to $z=0$.  The top panel shows results for the complete galaxy
sample, i.e., all galaxies above our $60 m_{\rm gas}=5\times 10^{10}M_\odot$ 
resolution threshold at the corresponding redshift, corresponding to
the right hand column of Figure~\ref{fig:evol}.
As noted in \S\ref{sec:gal}, there are
420, 1034, 2011, and 2571 galaxies satisfying this criterion at
$z=3$, 2, 1, and 0, respectively.  Since the strength of galaxy
clustering depends on mass (see \S\ref{sec:corrsub} below), it is
useful to identify samples of specified space density for comparison
to observations.  The middle panel shows results for the 500 most massive
galaxies present at each redshift (except at $z=3$, where it contains
only the 420 galaxies above the resolution threshold).  
The ranking of each galaxy is based on
its total mass of stars and cold ($T<30,000\;$K) gas.  
The comoving space density of this sample, 
$500/(50\hmpc)^3 = 0.004 h^3\mpc^{-3}$, is similar to that of 
$L_*$ galaxies today \citep{blanton01} and similar to (but slightly
higher than) the space density of Lyman-break galaxies in the
spectroscopic samples of \cite{adelberger98}.  We therefore refer to this as 
our ``$L_*$-sample,'' illustrated by the middle column of 
Figure~\ref{fig:evol}. 
At $z=0$, the baryon mass threshold for this sample
is $M_b=2.9\times 10^{11}M_\odot$.
The bottom panel of Figure~\ref{fig:corrall}
shows the dark matter $\xi(r)$.

In each panel of Figure~\ref{fig:corrall}, 
we show as a reference the power-law fit to the SDSS 
correlation function found by \cite{zehavi02}.  
Comparison of our predictions to results from a flux-selected observational 
sample rests on the implicit assumption that the luminosity of a galaxy is an
approximately monotonic function of its mass, and we use the SDSS
result in preference to the 2dFGRS result of \cite{norberg01}
because this assumption seems more applicable to the $r$-band
selection used in the SDSS.  Lyman-break galaxy samples at $z\approx 3$
are selected in the rest-frame UV, where the instantaneous star formation 
rate is more important than the stellar mass in determining galaxy
luminosity; however, in our simulations, these star formation rates
are fairly well correlated (though not perfectly so) with baryon mass
\citep{weinberg02}.

The main source of statistical error in our predictions is the finite
number of coherent structures present in our $(50\hmpc)^3$ simulation
volume.  We estimate the error bars on the correlation function and on
other statistical quantities that we compute later by
applying jackknife resampling to the eight octants of the simulation cube.
Specifically, we create eight jackknife subsamples by deleting each
of the eight octants of the cube in turn, and we estimate $\xi(r)$ for
each subsample; to avoid boundary effects that would complicate $\xi(r)$
estimation and reduce our effective volume, we identify pairs using the
full periodic cube and assign each pair to an octant based on the
position of one of its particles.  The $1-\sigma$ error bar $\sigma_i$
assigned to the value of $\xi(r)$ at $r=r_i$ is 
\begin{equation}
\label{eqn:jack}
\sigma_i^2 = {N \over N-1} \sum_{j=1}^{N} [\xi^j(r_i)-{\hat\xi}(r_i)]^2 ~,
\end{equation}
where $\xi^j$ is the estimate in subsample $j$, $\hat\xi$ is 
the estimate from the full cube, and $N=8$.
This approach is similar to using the dispersion among the values of
$\xi(r)$ from the eight octants (in which case one would divide by
$8^{1/2}$ to get the error on the mean), but it is more robust because
each of the subsamples is closer in size to the full sample.
\cite{zehavi02} use this method to estimate statistical errors on the
observed correlation functions of SDSS galaxies.
The jackknife error estimates automatically incorporate the contribution
from galaxy shot noise in addition to the finite number of structures,
but the latter usually dominates by a large factor over the former
(which would be represented by ``Poisson error bars''). 

The striking difference between the evolution of galaxy and dark matter
correlation functions in Figure~\ref{fig:corrall} is similar to that
found in previous studies based on hydrodynamic simulations,
semi-analytic calculations of galaxy formation, and high-resolution
N-body simulations that identify galaxies as ``sub-halos'' within
larger virialized objects 
\citep{katz99,colin99,kauffmann99b,cen00,benson01,pearce01,somerville01,
yoshikawa01}.
As Figure~\ref{fig:evol} shows,
galaxies (at least those massive enough to be resolved
by our simulation) form at special locations in the density field,
and at high redshift they already trace out the network of filaments
nascent in the dark matter distribution.  
The dark matter correlation function grows in time 
as mass moves into this network from the
surrounding regions, but the structure traced by galaxies stays
relatively unchanged, and the galaxy correlation function
is only weakly dependent on redshift.  

\begin{figure}
\centerline{
\epsfxsize=3.5truein
\epsfbox[120 265 460 720]{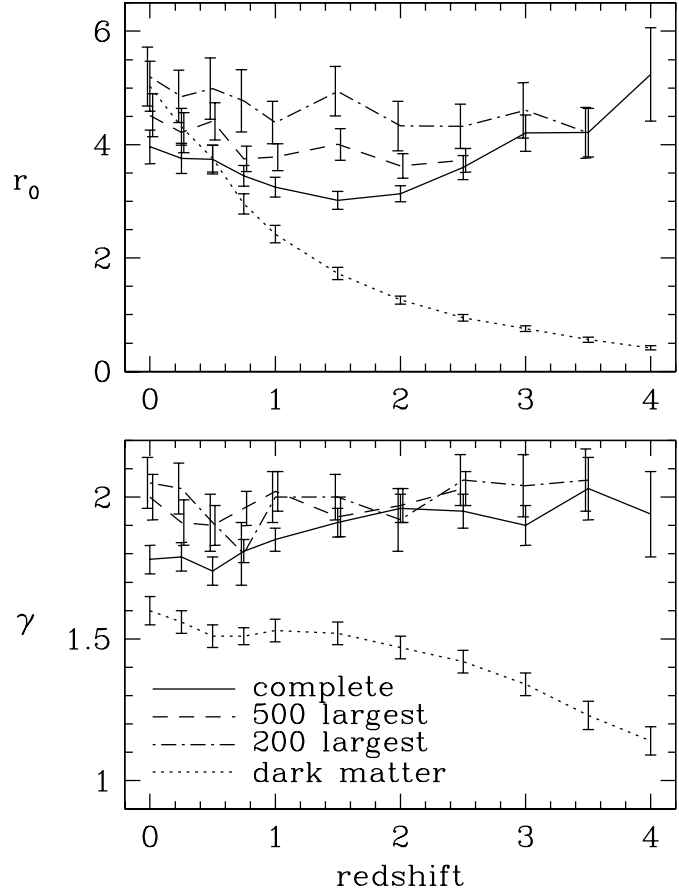}
}
\caption {Evolution of the correlation length $r_0$ (in comoving $\hmpc$)
and power-law index $\gamma$, for all galaxies (solid), the 500 most massive
galaxies (dashed), the 200 most massive galaxies (dot-dashed), and the
dark matter (dotted).  Error bars are obtained from the power-law fits,
using the jackknife errors on $\xi(r)$.  
Lines for the 500 largest galaxies stop at $z=2.5$, since the complete
sample contains fewer than 500 galaxies at higher redshift;
likewise, lines for the 200 largest galaxies stop at $z=3.5$.
}
\label{fig:r0gamma}
\end{figure}

\begin{figure}
\centerline{
\epsfxsize=3.5truein
\epsfbox[95 415 460 720]{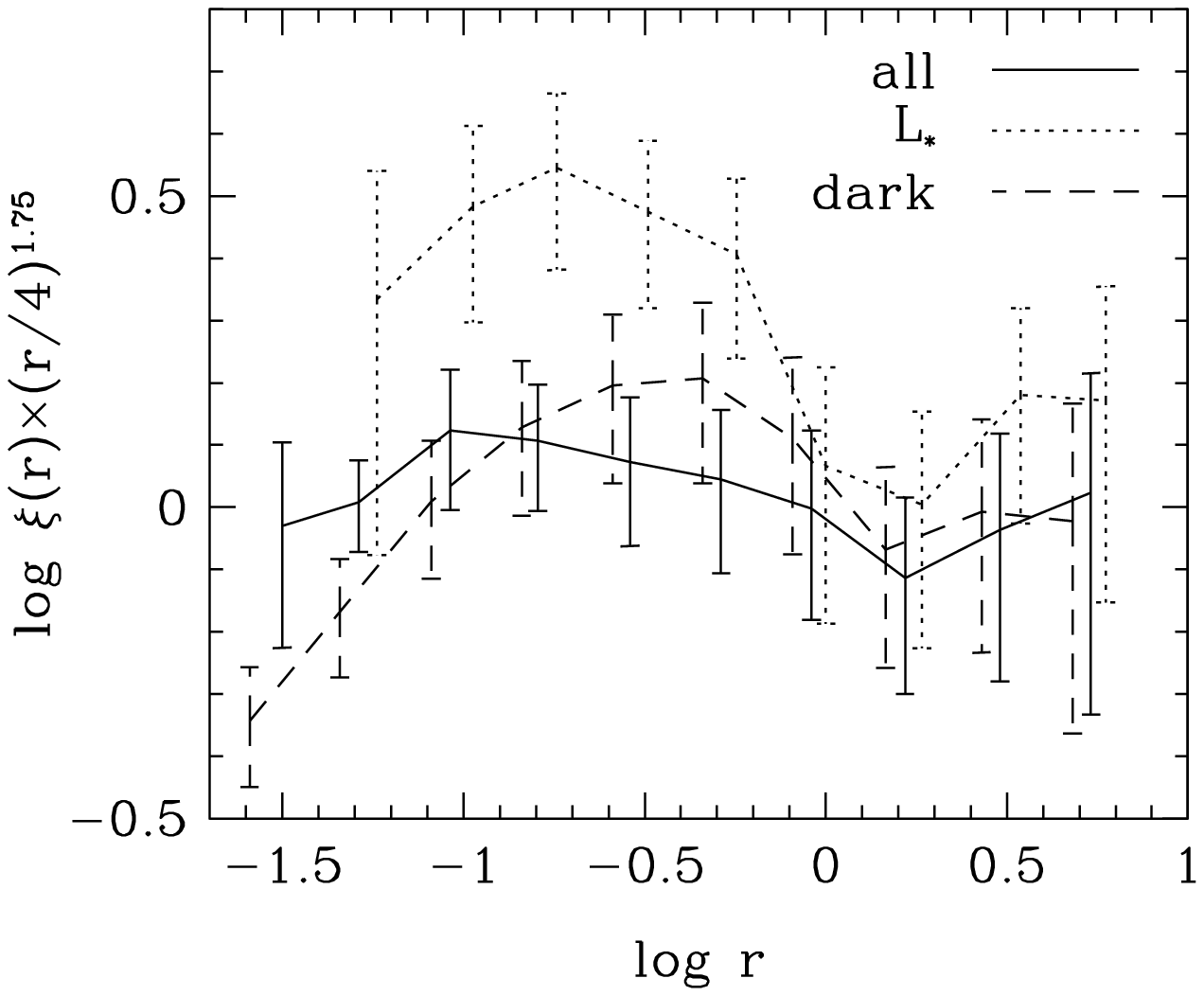}
}
\caption {
Ratio of the $z=0$ correlation functions to a power-law
$\xi(r)=(r/4.0\hmpc)^{-1.75}$, for the complete galaxy sample (solid),
the $L_*$ sample (dotted), and the dark matter (dashed).
Small horizontal offsets have been added to the jackknife error
bars to improve clarity.
On this plot, the SDSS correlation function would be a horizontal
line at $\log (6.1/4.0)^{1.75} = 0.32$, 
and the 2dFGRS correlation function of $L_*$ galaxies
would be a horizontal line at $\log (4.9/4.0)^{1.75} = 0.15$.
}
\label{fig:powerlaw}
\end{figure}

Figure~\ref{fig:r0gamma} quantifies these points in terms of the 
parameters $r_0$ and $\gamma$, which we determine from power-law
fits to the correlation functions over the full range plotted
in Figure~\ref{fig:corrall}, $0.02\hmpc < r < 12\hmpc$.
(Note that our figures plot the point for each radial separation bin
at the computed mean separation of pairs in that bin.)
For some sparse, high-redshift samples,
the smallest radial bins have no pairs, in which we case we start the
fit at slightly larger $r$.
We use the jackknife error bars on the $\xi(r)$ points to weight the
fit and determine its statistical errors.  We do {\it not} incorporate
the covariance of the $\xi(r)$ errors; in principle, we could estimate
the full error covariance matrix by the jackknife method, but in practice
such estimates are noisy and not easy to use.
Treating the $\xi(r)$ errors as independent causes us to underestimate 
the statistical uncertainty in $r_0$.
The dark matter correlation functions are not well described by
a power law at $z\la 1$, where the fits have $\chi^2/$d.o.f. of $\sim 1.5-2$,
so the derived values of $r_0$ and $\gamma$ are sensitive to the
radial range selected for the fit.  Power-law fits to the galaxy
correlation functions are nearly always adequate, and they usually
have $\chi^2/$d.o.f. less than one ($\sim 0.3-0.7$) because we have
ignored the covariance of the errors in $\xi(r)$.

Figure~\ref{fig:r0gamma} shows steady growth of the comoving 
correlation length of the dark matter, as expected.
The correlation length of the complete galaxy sample, on the other hand,
starts at $r_0=5.2\hmpc$ at $z=4$, declines to a minimum
of $3.0\hmpc$ at $z=1.5$, and climbs again to $r_0=4\hmpc$
by $z=0$.  The $L_*$ sample has a higher correlation length 
than the complete sample at $z=0$,
$r_0 \approx 4.5\hmpc$, though the correlation lengths of the two samples 
converge at higher redshift as the $L_*$ threshold approaches
our completeness threshold.
Correlation lengths for a sample with a higher mass threshold,
the 200 most massive galaxies at each redshift, are higher still,
$r_0=5.2\hmpc$ at $z=0$, with only a slight decline towards
higher redshift.
The values of $\gamma$ for the galaxy correlation functions range
from 1.75 to 2.05, with a tendency for larger $\gamma$ for the
more massive galaxy samples or at higher redshifts.
At high redshift, the dark matter correlation function is close
to a power law, but a quite shallow one.
As the scale of non-linear structure grows, it produces a ``bulge''
in the dark matter $\xi(r)$ (see Fig.~\ref{fig:corrall}), which drives the
correlation function away from a power law and steepens the
effective value of $\gamma$.

The slow evolution of the galaxy correlation length
 in Figure~\ref{fig:r0gamma} agrees with
results from the KPNO wide-area survey \citep{postman98}, which yields
$r_0\approx 4.5\hmpc$ (comoving) at $z\approx 0.5$, and from the
CNOC-2 redshift survey, which shows little evolution of the comoving
$r_0$ out to $z\sim 1$ \citep{carlberg99}.  However, the value of
$r_0$ depends significantly on galaxy mass and age 
(see Figure~\ref{fig:corrsub} below), so caution is required
when interpreting results of surveys that may select different types
of galaxies at different redshifts.  The predicted correlation
length at $z=3$, $r_0=4.2 \pm 0.3 \hmpc$, agrees very well with the value
$r_0=4.3\pm 0.3\hmpc$ measured from spectroscopic samples of 
Lyman-break galaxies (\citealt{adelberger02}; we have transformed from their
$\Omega_m=0.3$ cosmology to our $\Omega_m=0.4$ cosmology using the
ratio of angular diameter distances at $z=3$), though the slope
$\gamma = 1.90 \pm 0.07$ is steeper than the observed value
of $\gamma = 1.55 \pm 0.15$.  
Similar values of $r_0$ for Lyman-break galaxies have been obtained
using semi-analytic galaxy formation models 
\citep{governato98,kauffmann99b,benson01,somerville01} and
other hydrodynamic simulations \citep{cen00,pearce01,yoshikawa01}.
The value of $r_0$ from our $50\hmpc$ simulation is somewhat
larger than the estimate $r_0=3.1\hmpc$
(including a factor 2.2 correction for finite box size effects)
derived by \cite{katz99} from a simulation of an $11.111\hmpc$ cube,
but a difference of this sort is expected because the mass resolution
threshold of our present simulation is a factor of eight higher,
and correlation strength increases with galaxy mass.

Perhaps the most significant result of this section
is the power-law form of the galaxy
correlation function at $z=0$.  The dark matter correlation function
shows the curvature found in previous N-body studies
(e.g., \citealt{jenkins98}), and it is inconsistent with
a pure power law.  
We demonstrate this point explicitly in Figure~\ref{fig:powerlaw},
where we plot the ratio of the dark matter, complete galaxy,
and $L_*$-galaxy correlation functions to a power law $(r/4.0\hmpc)^{-1.75}$
that has the observed logarithmic slope.  The dark matter correlation 
function rises above this power law at $r\sim 0.5\hmpc$ and falls well
below it at $r \la 0.05 \hmpc$.  
(The departure from a power law is somewhat less prominent than that in
Jenkins et al.'s [\citeyear{jenkins98}] $\Lambda$CDM simulation 
because we adopt
$\sigma_8=0.8$ and $n=0.95$ instead of $\sigma_8=0.9$ and $n=1$.)
The correlation function of the
complete sample, on the other hand, follows the power law almost
perfectly.  The situation for the $L_*$ sample is less clear; here
$\xi(r)$ shows a rise at $r \la 1\hmpc$ reminiscent of that shown 
by the dark matter, but within our statistical errors, the $L_*$ $\xi(r)$
is adequately described by a power law with $\gamma = 2.00\pm 0.08$.

The correlation length of the $L_*$ sample, $r_0=4.5\pm 0.4\hmpc$, is smaller
than the SDSS value of $r_0=6.1\hmpc$, and slightly below the value
$r_0=4.9\hmpc$ found for blue-selected $L_*$ galaxies in the 2dFGRS.
The slope $\gamma=2.00\pm 0.08$ is steeper than the observed value
of $\gamma \approx 1.75$.  The low amplitude appears to be partly
an artifact of the particular realization of $\Lambda$CDM initial
conditions in our $50\hmpc$ simulation volume.  We have run particle-mesh
N-body simulations (using the code of Park [\citeyear{park90}]) of these
initial conditions and of four equivalent volumes with different random
realizations, and we find that the dark matter correlation length
of this realization is a factor of 1.2 lower than that derived from
the average correlation function of the other four realizations
($3.79\hmpc$ vs.\ $4.60\hmpc$), simply by chance.  There is also a 
systematic effect of missing power on scales larger than $50\hmpc$.
With PM simulations of $100\hmpc$ volumes, we find a further increase
in the mean $r_0$ by a factor of 1.1, and a slightly shallower slope
($\gamma=1.92$ over the range $1-10\hmpc$, vs. $\gamma=1.97$ for 
the $50\hmpc$ realization of the SPH simulation).  Unfortunately,
we do not know just how these differences in the dark matter $\xi(r)$
translate to the galaxy $\xi(r)$, but the statistical and systematic
effects in our $50\hmpc$ volume are large enough that we do not presently
regard the quantitative discrepancies between the predicted and observed
$(r_0,\gamma)$ for $L_*$ galaxies as significant.  The statistical
uncertainties can be reduced by using the halo occupation distribution
derived from the SPH simulation to populate a larger volume N-body
simulation, as discussed by \cite{berlind02b}.

\cite{berlind02} show that the key requirements for obtaining a power-law
$\xi(r)$ in a $\Lambda$CDM model are reducing the efficiency of galaxy
formation in the most massive (group and cluster scale) halos and
suppressing pair counts in the lowest mass halos by keeping fluctuations
about the mean occupation number well below the Poisson level
(see also \citealt{jing98,seljak00,peacock00,scoccimarro01}).  
The galaxy population
in our simulation satisfies both of these requirements, a point that
we discuss in detail elsewhere \citep{berlind02b}.
Our conclusion that including galaxy formation physics removes
most of the discrepancy between the predicted correlation function
of dark matter and the observed correlation function of galaxies agrees
with conclusions derived from other studies based on semi-analytic
modeling \citep{kauffmann99a,benson00a,benson00b,somerville01}, 
high resolution N-body simulations \citep{colin99},
and numerical hydrodynamics \citep{pearce99,pearce01,cen00,yoshikawa01}.

\subsection{Correlation Functions of Sub-populations}\label{sec:corrsub}

\begin{deluxetable}{lccl}
\footnotesize
\tablecaption{$r_0$ and $\gamma$ from simulated subsamples decomposed by age,
mass, and local galaxy density.  \label{tab:subsample} }
\tablewidth{0pt}
\tablehead{
\colhead{Subsample} & 
\colhead{$r_0$} & 
\colhead{$\gamma$}
} 
\startdata
All  & $3.96\pm 0.30$ & $1.78\pm 0.05$ \\
Old & $5.24\pm 0.47$ & $1.84\pm 0.06$ \\
Young & $2.77\pm 0.17$ & $1.82\pm 0.06$ \\
High mass & $4.00\pm 0.29$ & $1.90\pm 0.06$ \\
Low mass  & $3.95\pm 0.41$ & $1.57\pm 0.08$ \\
High density & $7.82\pm 0.71$ & $1.77\pm 0.06$ \\
Low density  & $1.89\pm 0.08$ & $2.07\pm 0.05$ \\
\enddata
\end{deluxetable}

\begin{figure}
\centerline{
\epsfysize=6.0truein
\epsfbox[125 75 460 720]{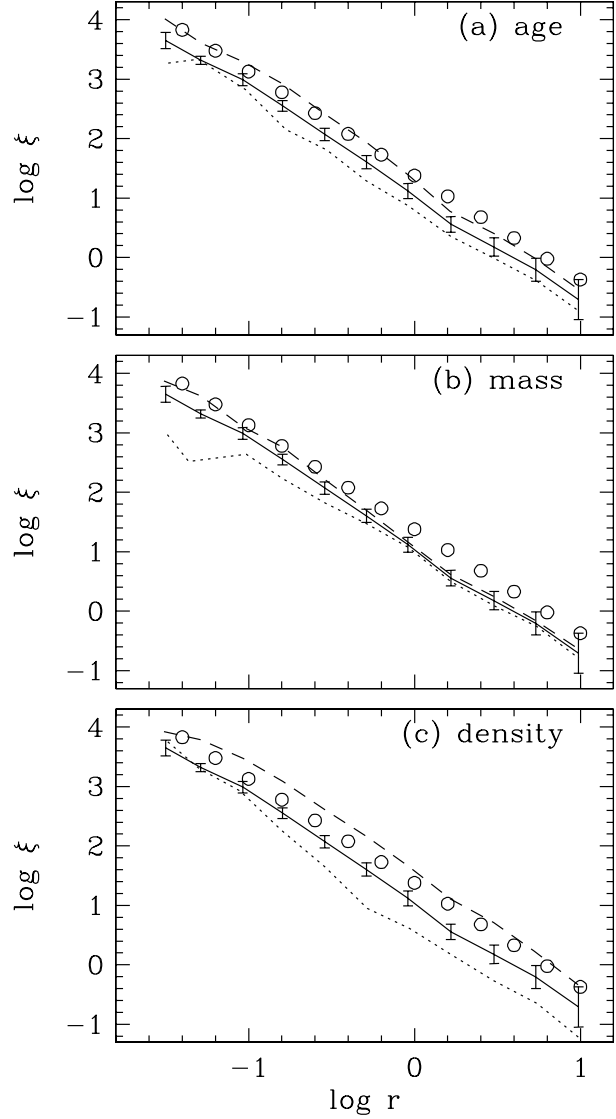}
}
\caption { Dependence of the galaxy correlation function on 
stellar population age (top),
baryon mass (middle), and density of the local environment
(bottom), at $z=0$.
In each panel, the solid line shows the correlation function and
jackknife error bars of the complete galaxy sample, and open
circles show the power-law fit to the SDSS correlation function.
Dashed lines show the correlation function of the older,
more massive, or higher density half of the complete sample
in panels (a)-(c), respectively, and dotted lines show the
correlation function of the other half of the sample.
}
\label{fig:corrsub}
\end{figure}

As mentioned in \S\ref{sec:corrall}, the galaxy correlation function 
is known to vary significantly with galaxy luminosity, color, and
morphology (e.g., \citealt{davis76,hamilton88,loveday95,guzzo97,willmer98}).
The most comprehensive recent examinations
are the studies of luminosity and color dependence for SDSS galaxies
by \cite{zehavi02} and the studies of luminosity and spectral type
dependence for 2dFGRS galaxies by \cite{norberg01,norberg02b}.
\cite{zehavi02} find a steady increase of correlation amplitude with 
luminosity, going from $\sim 1.5$ magnitudes fainter than $M_*$ to $\sim 1.5$
magnitudes brighter, and they find that red galaxies have a higher
amplitude and steeper correlation function than blue galaxies, in
agreement with earlier studies of the difference between clustering
of early type (red) and late type (blue) samples.
\cite{norberg01} find a somewhat different luminosity dependence
for the $b_{\rm J}$-selected 2dFGRS galaxies, with little
dependence of clustering amplitude on luminosity below $L_*$ but
a strong dependence at higher luminosities.  \cite{norberg02b}
find that the correlation amplitude varies strongly with galaxy
spectral type, analogous to the usual dependence on morphological type,
and that the luminosity dependence is present within each spectral type
class.  Galaxies with early spectral types have a somewhat steeper
correlation function than those with late spectral types, though
the difference in slope is not as striking as that for the SDSS red
and blue galaxy samples.

Our simulation volume is too small and our computation of galaxy 
properties too simple to allow detailed comparisons to these
observations, but we can examine qualitative trends.
The color and spectral type of a galaxy depends largely on the age of its 
stellar population, and morphology is also known to correlate strongly with
population age.
For our simulated galaxy population, we define a galaxy's age by the 
epoch at which it formed half of its stars.
While this approach necessarily brings our heuristic star formation 
algorithm into play, the details of this algorithm should have little 
effect on the conclusions presented here, since we will use only
the {\it relative} ages of the simulated galaxies.

Figure~\ref{fig:corrsub}a shows the correlation functions of the
older (dashed line) and younger (dotted line) halves of the complete
galaxy sample, compared to the correlation function of the complete
sample (solid line).  The correlation amplitude exhibits a clear
dependence on galaxy age; the older galaxies have a correlation 
length of $r_0=5.2\hmpc$ and the younger galaxies have $r_0=2.8\hmpc$.
The fitted slopes of the correlation functions are nearly identical,
$\gamma=1.84$ and 1.82, respectively.
(Values of $r_0$ and $\gamma$ for all sub-populations appear
in Table~\ref{tab:subsample}.)
These results are in good qualitative agreement with 
the spectral type dependence found by 2dFGRS; relative to the
SDSS results, they reproduce the observed color dependence
of correlation amplitude but not the observed change of slope.

Figure~\ref{fig:corrsub}b compares the correlation functions of the
more massive (dashed) and less massive (dotted) halves of the
complete sample.  While the more massive galaxies have a higher
$\xi(r)$ at all separations, the difference is marked only 
below $\sim 0.5\hmpc$.  Power-law fits yield nearly identical
correlation lengths, $r_0 \approx 4.0\hmpc$, but significantly
different slopes ($\gamma=1.90$ vs. $\gamma=1.57$).  
Table~\ref{tab:r0gamma} shows that $r_0$ does rise significantly
if one takes the 500 most massive ($r_0=4.5\hmpc$) or 200 most
massive ($r_0=5.2\hmpc$) galaxies, and in combination with 
Figure~\ref{fig:corrsub}b it suggests that this mass dependence
sets in at roughly the space density of $L_*$ galaxies.
The 2dFGRS results show a transition of this sort
\citep{norberg01}, though the SDSS results show a steadier dependence
of clustering amplitude on luminosity \citep{zehavi02}.

Figure~\ref{fig:corrsub}c shows the effect of dividing the galaxy
sample based on local density, computed by smoothing with a spline
kernel (the same form used in the SPH calculation) whose radius is adjusted 
to always enclose 16 galaxies within the smoothing volume.
Since $\xi(r)$ is itself a measure of overdensity at radius $r$,
it is no surprise that this division produces the most marked
change in the correlation function.  The enhancement of $\xi(r)$
persists to scales much larger than the typical smoothing radius
because of the bias of locally dense
regions towards overdense large scale environments 
(\citealt{kaiser84}; see \citealt{mann98} and \citealt{narayanan00}
for discussions in the context of N-body models).
The observed dependence of galaxy morphology on local density
(e.g., \citealt{dressler80,postman84}) should therefore lead to
a large scale bias in the clustering of early-type galaxies.

To summarize the results of this and the previous section, we find
that our simulated galaxy population reproduces many of the features
of observed galaxy correlation functions: a power-law form with 
$\gamma\approx 1.75$ down to separations $r\sim 0.02\hmpc$, an increase
of correlation strength with mass that becomes more pronounced
at high masses, and a higher correlation amplitude for galaxies
with older stellar populations.  The trends of clustering strength
with galaxy mass and age agree with the predictions from 
semi-analytic models \citep{kauffmann99a,benson00a,somerville01} 
and other hydrodynamic
simulations \citep{cen00,pearce01,yoshikawa01}.  
The predicted correlation lengths are somewhat low and the $\xi(r)$
slope for $L_*$ galaxies slightly too steep in comparison to the 
SDSS results, but our N-body experiments suggest that these
discrepancies are largely explained by the truncation of power
at the scale of our $50\hmpc$ box and by a lower than average
clustering amplitude in the particular $\Lambda$CDM initial
conditions used in our simulation.

\subsection{Pairwise Velocities}\label{sec: vel}

\begin{figure}
\centerline{
\epsfxsize=5.0truein
\epsfbox[55 220 550 720]{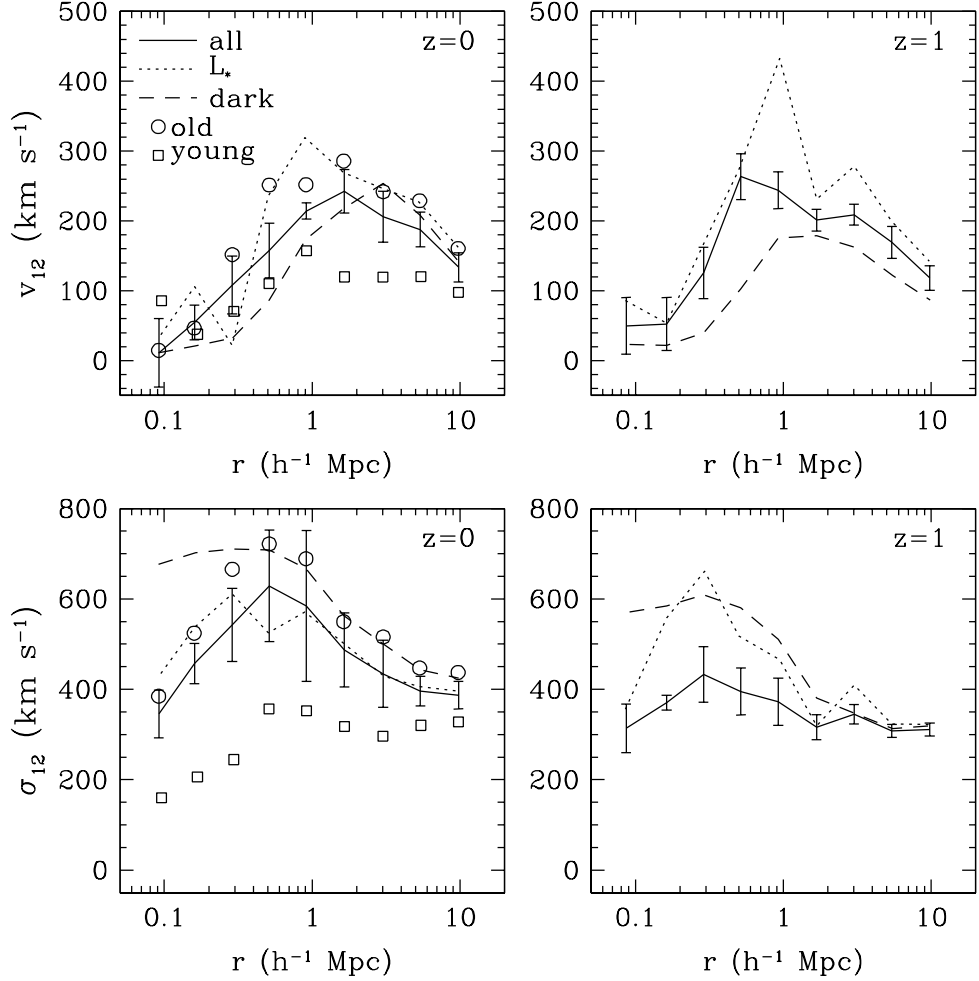}
}
\caption {
Mean pairwise radial velocities (top panels) and pairwise radial 
velocity dispersions (bottom panels) at $z=0$ and $z=1$, as
indicated.  Solid lines with jackknife error bars show results
for the complete galaxy sample, while dotted and dashed lines show
results for the $L_*$ sample and the dark matter, respectively.
In the $z=0$ panels, circles and squares show results for the
older and younger halves of the complete sample, respectively.
The pairwise velocity is defined with Hubble flow removed and
a convention of positive sign for galaxies moving towards each other.
The pairwise dispersion is 1-dimensional, along the direction of
separation, with the mean pairwise velocity subtracted.
}
\label{fig:pv}
\end{figure}

\cite{peebles80}, \cite{davis83}, and \cite{bean83} describe the
techniques now commonly used in correlation function analyses
of galaxy redshift surveys.  Starting from $\xi(r_p,\pi)$,
the correlation as a function of projected and line-of-sight
separations, one integrates over $\pi$ to obtain the projected
correlation function $w(r_p)$, which can be easily inverted
to obtain the real space 3-d correlation function $\xi(r)$.
The full $\xi(r_p,\pi)$ can be calculated, approximately,
by a convolution of $\xi(r)$ with the galaxy pairwise velocity
distribution (see Fisher [\citeyear{fisher95}] for a careful
discussion of this approximation and its relation to the linear
theory approximation of Kaiser [\citeyear{kaiser87}] and
Hamilton [\citeyear{hamilton92}]).  This distribution is specified
by its first and second moments $\vpair$ and $\sigma^2_{12}(r)$,
and by a functional form, usually assumed (on both theoretical
and observational grounds) to be exponential.  The pairwise
dispersion can be inferred from $\xi(r_p,\pi)$ in a fairly
robust fashion, though it is sensitive to the presence of rare,
rich clusters in the survey volume \citep{mo93}.
The mean pairwise velocity can in principle be deduced as well,
given sufficiently good data (see, for example, the analysis of
\citealt{fisher94}); at larger separations, it can also be measured
directly in galaxy peculiar velocity surveys \citep{ferreira99}.

Figure~\ref{fig:pv} shows the mean pairwise velocity (top)
and pairwise velocity dispersion (bottom) at $z=0$ (left) and
$z=1$ (right).  At $z=1$, the mean pairwise velocity of the
galaxies is higher than that of the dark matter, at all separations
$0.1-10\hmpc$.  Although galaxies and dark matter feel the
same gravitational forces and therefore have similar large scale
velocity fields, pair-weighted statistics like $\vpair$ and $\spair$
are also sensitive to spatial bias; in this case, the preferential
location of galaxies in overdense regions with high inflow
velocities boosts $\vpair$. The mean
pairwise velocities of the $L_*$ sample are higher than those of the
complete sample, but the difference is of marginal statistical significance.

By $z=0$ (top left), the bias of the galaxy mean pairwise velocities has
largely disappeared, in correspondence with the small spatial bias
at $z=0$ (see Figure~\ref{fig:corrall}, and Figure~\ref{fig:bsigma}
below).  The complete and $L_*$ galaxy samples again exhibit 
similar $\vpair$, with velocities of the $L_*$ sample being
slightly higher.  There is a much sharper difference, however, between
the older and younger halves of the complete sample
(circles and squares in Figure~\ref{fig:pv}), with the older
galaxies having nearly twice the value of $\vpair$ on scales of
one to several $\hmpc$.  This difference reflects the preference
of older galaxies for denser environments with larger inflow velocities.

The pairwise dispersion of the complete galaxy sample is below that of
the dark matter both at $z=1$ and at $z=0$.
The $L_*$ sample's pairwise dispersion is similar to that of the
dark matter at $z=1$ and below it at $z=0$.  All of the 
galaxy samples show a peak in $\spair$ at $r\sim 0.3\hmpc$,
roughly the half-mass radius of the richest groups.
While one might expect suppression of the galaxy pairwise dispersion 
to go hand-in-hand with spatial anti-bias, our results show
that the story is not so simple.  For example,
the correlation function of the $L_*$ sample at $z=0$
is positively biased at all separations (Figure~\ref{fig:powerlaw}),
but its pairwise dispersion is lower than that of the dark matter
and nearly identical to that of the (more weakly clustered) complete
galaxy sample.  As emphasized by \cite{berlind02} and
\cite{sheth01}, the pairwise dispersion is influenced by factors
that do not affect $\xi(r)$ (or even affect it with opposite sign),
making it sensitive to details of the relation between galaxies
and their surrounding dark matter halos.  
While $\spair$ is not sensitive
to galaxy baryon mass (at least in the comparison presented here), 
it is highly sensitive to stellar population age: 
the older half of the complete sample has a substantially
higher pairwise dispersion than the younger half because of the
location of older galaxies in more massive virialized structures.

The pairwise dispersion of the complete and $L_*$ galaxy samples agrees 
reasonably well with the values $\sigma_{12} \sim 550-650\kms$ at
$r \sim 1\hmpc$ found in the CfA2, LCRS, and SDSS redshift surveys
\citep{marzke95,jing98,zehavi02}.  However, the sensitivity of $\spair$
to rare, rich clusters is reflected in the large jackknife error
bars of Figure~\ref{fig:pv}, so we cannot take this agreement
as a major triumph of the simulation.  The scale-dependence in
Figure~\ref{fig:pv} contrasts with the rather flat $\spair$ 
inferred from the observations, but for the same reason we do not
view this as a serious discrepancy.  We can conclude from 
Figure~\ref{fig:pv} that the predicted difference between dark matter
and galaxy pairwise dispersions makes it easier to understand the 
relatively low observed value of $\sigma_{12}$, and that the predicted
difference between old and young galaxies naturally accounts 
for the factor of $\sim 2$ difference in $\sigma_{12}$ derived
for red and blue galaxies in the SDSS \citep{zehavi02} and the factor
$\sim 1.7$ difference in $\sigma_{12}$ derived from optical
redshift surveys (as cited above) and from the IRAS PSCz catalog
(as analyzed by \citealt{jing02}).

Our conclusion that the galaxy pairwise dispersion at $z=0$ is significantly
lower than that of the dark matter agrees with the results of 
\cite{pearce01}, based on a hydrodynamic simulation, and \cite{benson00a},
based on populating N-body halos according to semi-analytic prescriptions.
We also agree fairly well with these calculations in terms of the
amplitude of the pairwise dispersion, $\sigma_{12} \sim 500-600\kms$
at $r \sim 1\hmpc$, and the shape of $\spair$, though our pairwise 
dispersion drops somewhat more rapidly at $r < 0.5\hmpc$.
\cite{kauffmann99a}, also using N-body halos and semi-analytic galaxy
models, find only a slight difference between the dark matter and
galaxy pairwise dispersions, and their predicted amplitude at $r\sim 1\hmpc$
is $\sigma_{12} \sim 800\kms$.  \cite{benson00a} argue that the difference
between their results and Kauffmann et al.'s arises because their 
(Benson et al.'s)
semi-analytic model places fewer galaxies in high dispersion
halos.  Since the halo occupation of our simulation agrees well with
that of the \cite{benson00a} models \citep{berlind02b},
we think that the same explanation applies to our results.
We agree with \cite{kauffmann99a} with respect to the mean pairwise
velocity (not examined by the other groups), with both calculations 
showing little difference between the galaxies and the dark matter at $z=0$.
While age divisions (or color divisions) were not examined in these
other studies, the tendency to find older, redder galaxies in more massive
halos appears to be a generic prediction of both hydrodynamic and
semi-analytic calculations, so we expect the strong age-dependence of
$\spair$ and $\vpair$ to be a fairly generic result.

\subsection{Moments of Galaxy Counts}

\begin{figure}
\centerline{
\epsfxsize=3.5truein
\epsfbox[105 415 460 720]{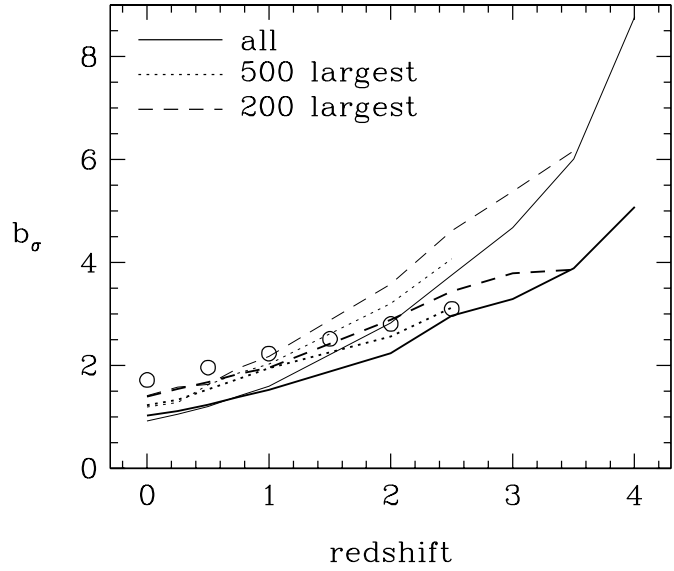}
}
\caption {
Evolution of the bias factor $b_\sigma$, defined by the ratio
of rms galaxy count fluctuations (corrected for shot noise) to
rms dark matter fluctuations in spheres of comoving radius $8\hmpc$
(heavy lines) or $2\hmpc$ (light lines).
Solid, dotted, and dashed lines show results for the complete galaxy
sample, the 500 most massive galaxies, and the 200 most massive
galaxies, respectively.
Lines for the 500 largest galaxies stop at $z=2.5$, since the complete
sample contains fewer than 500 galaxies at higher redshift;
likewise, lines for the 200 largest galaxies stop at $z=3.5$.
Circles show the number-conserving model of \cite{fry96}, normalized to a bias
factor of $3.1$ at redshift $z=2.5$.
}
\label{fig:bsigma}
\end{figure}

\begin{figure}
\centerline{
\epsfxsize=3.5truein
\epsfbox[105 425 460 720]{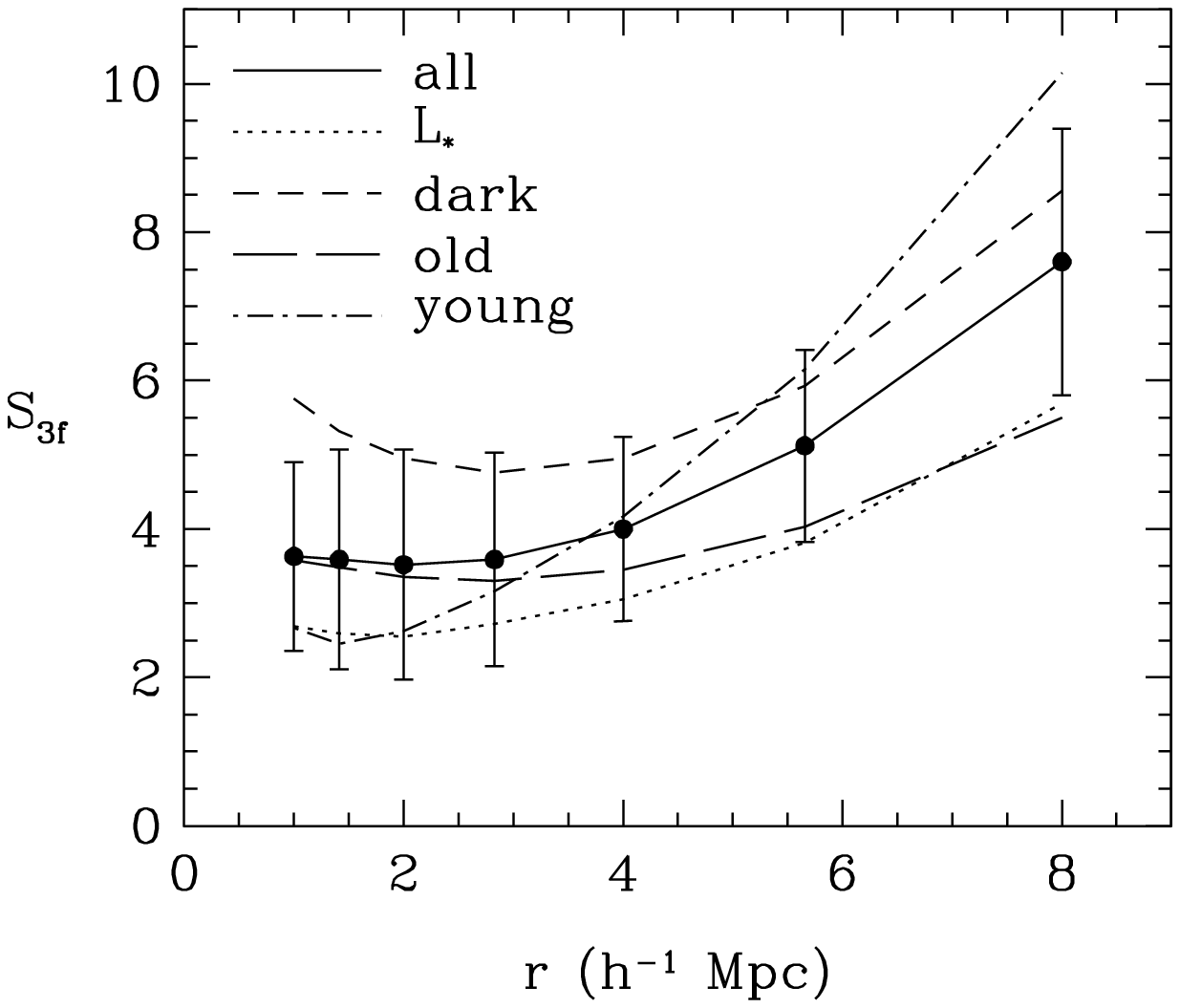}
}
\caption {
The hierarchical amplitude $\stf$, defined as the
ratio of the third factorial moment to the square of the second 
factorial moment in spheres of radius $r$, at $z=0$.
Points connected by the solid line show results for the 
complete galaxy sample, with jackknife error bars.
Other lines show results for the dark matter (short-dashed),
the $L_*$-sample (dotted), and the older (long-dashed) and
younger (dot-dashed) halves of the complete sample.
The value of $S_{3}$ inferred from angular clustering in
the APM galaxy catalog is roughly constant at $S_3\approx 3.5$
over this range of scales.
}
\label{fig:s3}
\end{figure}

Moments of the galaxy count distribution in spherical (or other)
cells provide an alternative to correlation functions for characterizing
galaxy clustering.  The $n$-th central moment of the galaxy counts
is closely related to the $n$-point correlation function averaged
over the cell volume.  As pointed out by \cite{szapudi93}, for a 
discrete galaxy distribution it is convenient to work with
factorial moments such as $\langle N(N-1) \rangle$ and
$\langle N(N-1)(N-2) \rangle$ (where $N$ is the number of galaxies
within a cell or sphere), since these automatically remove
Poisson shot noise contributions and can therefore be related
to the moments of an underlying continuum distribution in a 
straightforward way.

The ratio $b_\sigma$ of rms galaxy count fluctuations to rms
mass fluctuations provides a simple and robust measure of 
galaxy bias, and the value of $b_\sigma$ for spheres of radius $8\hmpc$
is the number most often quoted as the ``bias factor'' of the
galaxy distribution.  Figure~\ref{fig:bsigma} shows $b_\sigma(8\hmpc)$
and $b_\sigma(2\hmpc)$ as a function of redshift, with 
\begin{equation}
b_\sigma \equiv 
  \left(\frac{\langle N(N-1)\rangle}{\langle N \rangle^2} -1 \right)^{1/2}
  \times
  \left\langle\left(\frac{\delta\rho}{\rho}\right)^2\right\rangle^{-1/2}
\label{eqn:bsigma}
\end{equation}
and the moments evaluated using $10^6$ spheres placed randomly in the
simulation volume.  At $z=0$ the bias factors are similar for $r=2\hmpc$
and $8\hmpc$, with $b_\sigma(8\hmpc)=1.02$ for the complete galaxy sample,
1.22 for the $L_*$ sample, and 1.40 for the 200 most massive galaxies.
Figures~\ref{fig:r0gamma}--\ref{fig:pv} show that the simulated galaxies
do not trace the mass in detail and that galaxy clustering depends on age
and luminosity, but our complete galaxy sample is, in an rms sense,
nearly unbiased at $z=0$.

As shown earlier in terms of $\xi(r)$, the galaxy bias starts high
and declines with redshift as the dark matter clustering catches up.
For the complete sample, $b_\sigma(8\hmpc)=3.3$ at $z=3$.
At redshifts $z\sim 1$, the bias factor is significantly larger
at $2\hmpc$ than at $8\hmpc$, presumably reflecting the different
behavior of $b_\sigma$ in the non-linear and linear regimes.
\cite{fry96} discusses an analytic model in which the galaxy population
is born at some redshift $z_*$ with a bias factor $b_*$ but thereafter
moves with the same velocity as the dark matter and conserves number
(see also \citealt{tegmark98}).  In this case the bias factor at
redshift $z<z_*$ is given by $b(z)-1 = (b_*-1)D(z_*)/D(z)$, where
$D(z)$ is the linear growth factor at redshift $z$.  Open circles in
Figure~\ref{fig:bsigma} show the prediction of this model, normalized
to the bias factor $b_*=3.1$ of the 500-galaxy sample at $z_*=2.5$.
The actual bias factor of this sample declines significantly faster
than the number-conserving model predicts, even though the sample
contains 500 galaxies at all redshifts, presumably because mergers
drive down the number of pairs in dense regions and galaxies that
form at $z<2.5$ are born in less biased environments.
\cite{somerville01} obtain a similar result with semi-analytic modeling.

Discussions of the third moment of the count distribution often focus
on the hierarchical ratio 
$S_3\equiv \langle \delta^3\rangle / \langle \delta^2 \rangle^2$.
For spherical cells and a linear power spectrum $P(k)\propto k^n$,
the dark matter has $S_3 = 34/7 - (3+n)$ to second order
in perturbation theory, independent of sphere radius and power
spectrum amplitude and nearly independent of $\Omega_m$
\citep{juszkiewicz93}.  If the galaxy density contrast is a local
function of the matter density contrast, $\delta_g=f(\delta)$,
then the relation between $S_{3g}$ and $S_3$ (representing the
galaxy and mass moments, respectively) depends on both the linear
bias factor $b_\sigma$ and a second bias factor $b_2$ that 
characterizes the second derivative $f^{\prime\prime}(\delta)$
at $\delta=0$ \citep{fry93,juszkiewicz95}.  The behavior of $S_3$
and the relation between $S_{3g}$ and $S_3$ becomes more complicated
in the fully non-linear regime where second-order perturbation
theory breaks down, but it remains the case that $S_3$ is only weakly
dependent on scale and that the ratio $S_{3g}/S_3$ depends on 
the non-linearity of the relation between galaxy and mass densities.

Figure~\ref{fig:s3} shows the hierarchical ratio $\stf$ defined in terms
of factorial moments,
\begin{equation}
\stf \equiv
  \left(\frac{\langle N(N-1)(N-2)\rangle}{\langle N \rangle^3} -1 \right)
  \left(\frac{\langle N(N-1)\rangle}{\langle N \rangle^2} -1 \right)^{-2},
\label{eqn:s3f}
\end{equation}
at $z=0$.  For the complete galaxy sample, $\stf \approx 3.5-4$
for $r=1-4\hmpc$, rising to $\stf=6.5$ at $r=8\hmpc$.
These values are slightly below those of the dark matter distribution.
Ratios for the more massive, $L_*$ sample are lower still; these
galaxies are more clustered in an rms sense, but they are less ``skewed''
in the sense measured by $\stf$.  As emphasized by \cite{colombi00},
the statistical and systematic errors in measurements of $S_3$ from 
a volume this size are substantial, and our jackknife error bars are
large.  Thus, while the predicted value of $\stf$ for the complete
galaxy sample agrees well with the value $S_3\approx 3.5$ that
\cite{gaztanaga94} infer by de-projecting the APM angular count moments, 
we cannot claim this as a major success.  (The recent analyses of
angular clustering in the SDSS by Gazta\~naga [\citeyear{gaztanaga02}]
and Szapudi et al.\ [\citeyear{szapudi02}] suggest that the APM 
numbers should be adjusted upward by $\sim 20\%$.)
However, the relative values of $\stf$ for the different populations
should be more robustly predicted, and our simulation indicates
that galaxies comparable to those in the complete sample should
have a hierarchical ratio slightly below that of the dark matter,
and that $\stf$ should be smaller for more massive galaxies.
The most striking result in Figure~\ref{fig:s3} is the strong
scale-dependence of $\stf$ for young galaxies, a prediction
that should be testable with the SDSS and 2dFGRS surveys in the
near future.

\section{Galaxy-Mass Correlations}\label{sec: gdm}

Advances in wide-field CCD imaging have opened a new window on the relation
between galaxies and mass: weak lensing measurements of the galaxy-mass
cross-correlation function.  This statistic can be derived from galaxy-galaxy
lensing analyses, which measure the mean shear profile around a sample of
foreground galaxies using the tangential distortion of background galaxies
\citep{tyson84,brainerd96,dellantonio96,griffiths96,hudson98,fischer00,
hoekstra01,mckay01,smith01,wilson01b}.
Alternatively, one can correlate the mass map derived from the ellipticity
correlation function of faint galaxies with the light map of brighter,
foreground galaxies in the same images \citep{wilson01a}.  In addition,
cosmic shear measurements are beginning to yield measurements of
the matter correlation function itself (for specified $\Omega_m$),
with impressively good agreement among independent surveys
(see \citealt{mellier01}; \citealt{hoekstra02}; and references therein),
allowing for the comparison of galaxy and mass fluctuation amplitudes.

\begin{figure}
\centerline{
\epsfysize=6.0truein
\epsfbox[125 75 460 720]{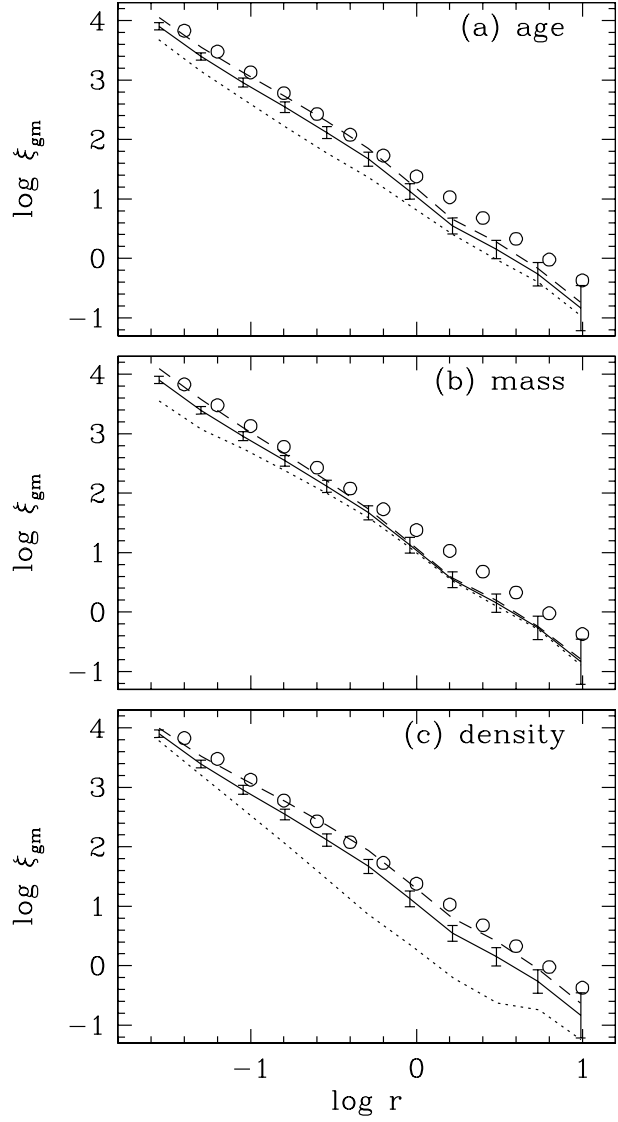}
}
\caption { 
The cross-correlation function $\xi_{\rm gm}(r)$ 
of galaxies and mass (dark matter + baryons) at $z=0$.
In each panel, the solid line with jackknife error bars
shows results for the complete galaxy sample.
Dashed lines show results for the older, more massive,
or higher density half of the complete sample in panels
(a)-(c), respectively, and dotted lines show results
for the other half of the sample.
Circles are the same as those in Fig.~\ref{fig:corrsub},
to facilitate visual comparison to the autocorrelation functions.
}
\label{fig:gmxisub}
\end{figure}

\begin{figure}
\centerline{
\epsfxsize=3.5truein
\epsfbox[50 50 530 720]{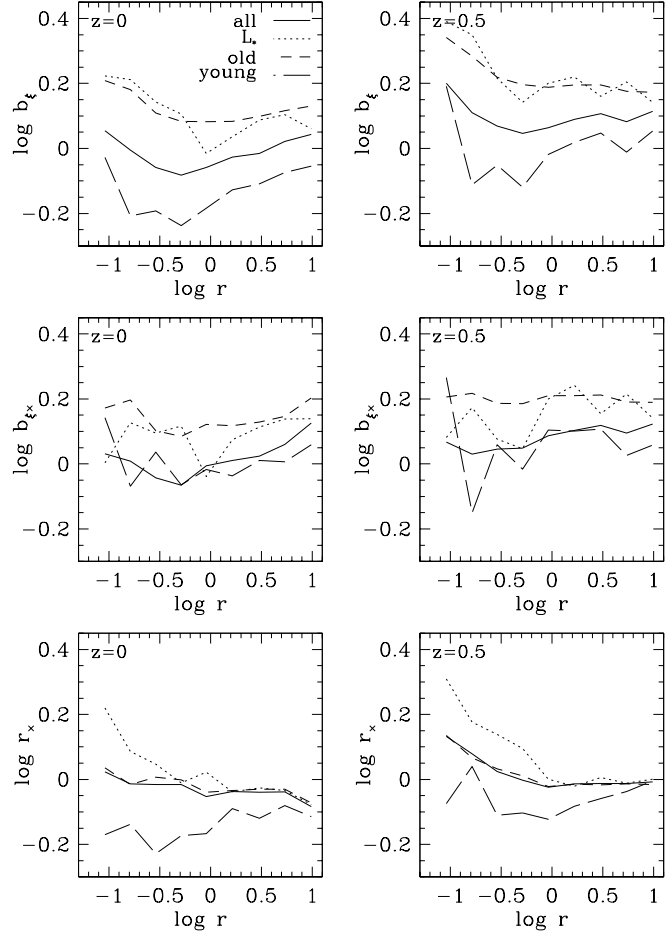}
}
\caption { 
Scale-dependence of the bias factors defined by 
$\bxi(r)=[\xi_{\rm gg}(r)/\xi_{\rm mm}(r)]^{1/2}$ (top) and by
$\bxix(r)=\xi_{\rm gg}(r)/\xi_{\rm gm}(r)$ (middle), 
and the ``correlation coefficient''
$\rx(r)=\bxi(r)/\bxix(r)$ (bottom), where
$\xi_{\rm gg}(r)$ and $\xi_{\rm mm}(r)$ are the galaxy and mass
autocorrelation functions and $\xi_{\rm gm}(r)$ is the 
galaxy-mass cross-correlation function.
Results are shown for the complete galaxy sample (solid),
the $L_*$-sample (dotted), and the older (short-dashed)
and younger (long-dashed) halves of the complete sample,
at redshifts $z=0$ (left column) and $z=0.5$ (right column).
}
\label{fig:gmxibias}
\end{figure}

Figure~\ref{fig:gmxisub} shows the galaxy-mass cross correlation function
for the complete galaxy sample and for the subsamples divided by age (top),
mass (middle), and local density (bottom).  We define 
$\xigm(r) = \langle \delta_g(\vx) \delta_m(\vx+\vecr)\rangle$, where
$r=|\vecr|$ and $\delta_g = n_g/{\bar n}_g-1$ and $\delta = \rho/{\bar \rho}-1$
are the galaxy and mass (dark + baryonic) overdensities; we compute it by
counting galaxy-particle pairs, with particles weighted by mass.
While the form of $\xigm(r)$ at small $r$ is connected to the typical
galaxy halo profile, the connection is rather loose because the galaxies
are not isolated and the samples contain objects with a range of halo
virial radii (points emphasized by \cite{guzik01}, \cite{white01},
and \cite{yang02}).
We use the same format in Figure~\ref{fig:gmxisub} and 
Figure~\ref{fig:corrsub}, and the two figures bear a striking resemblance
to each other.  To a first approximation, the galaxy-mass correlation
function has the same shape as the galaxy-galaxy correlation function, similar
amplitude, and similar dependence on galaxy age, baryon mass, and 
environment.  However, the variations of $\xigm(r)$ for different 
subsamples are smaller than the variations of $\xigg(r)$, as one would 
expect, given that the mass distribution stays the same regardless of the
galaxy subsample.  

With measurements of $\xigg(r)$, $\xigm(r)$, and the mass correlation
function $\ximm(r)$, one can define two ``bias functions,''
$\bxi(r) \equiv [\xigg(r)/\ximm(r)]^{1/2}$ and
$\bxix(r) \equiv \xigg(r)/\xigm(r).$
The ratio $\rx(r) \equiv \bxi(r)/\bxix(r) = 
\xigm(r)/[\xigg(r)\ximm(r)]^{1/2}$ corresponds to a
``correlation coefficient'' as defined by, e.g., \cite{tegmark99},
though here we work in terms of correlation functions rather
than power spectra.
In the case of linear, deterministic bias, where 
$\delta_g(\vx) = b\delta(\vx)$ at all positions $\vx$,
$\bxi$ and $\bxix$
would be equal to $b$ and independent of scale, and $\rx$ would equal unity.
Figure~\ref{fig:gmxibias} plots $\bxi(r)$, $\bxix(r)$, and $\rx(r)$
for the complete galaxy sample, the $L_*$ sample, and the older and
younger halves of the complete sample.  
We will focus on the $z=0$ results, but because measurements of these
quantities come from deep imaging surveys, we also show the simulation
predictions at $z=0.5$, which are qualitatively similar.

The complete sample has $\bxi \approx 1$, but there is a broad
minimum around $0.5\hmpc$, with $\bxi(0.5\hmpc)\approx 0.75\bxi(10\hmpc)$.
It is precisely this mild scale-dependence of $\bxi(r)$ that transforms
the curved dark matter correlation function into a power-law galaxy
correlation function (see Figure~\ref{fig:powerlaw}).
The $L_*$ sample has higher $\bxi$, roughly flat (though noisy)
at $r\ga 0.5\hmpc$, and climbing towards small $r$.
The old galaxies have $\bxi(r)$ similar to that of the 
$L_*$ sample, while the young galaxies have low $\bxi(r)$ with
a scale-dependence similar to that of the complete sample.

The scale-dependence of $\bxix(r)$ is generally weaker than that of
$\bxi(r)$.  At $r\ga 1\hmpc$, values of $\bxix$ are similar to those
of $\bxi$ for the corresponding sample, except in the case of the
young galaxies, which have $\bxix>\bxi$ and thus $\rx(r)<1$.
This ``anti-correlation'' arises because the young galaxies tend to avoid
the higher mass halos, which contribute significantly to $\ximm(r)$.
The complete, $L_*$, and old galaxy samples have $\rx(r) \approx 1$
at large scales and $\rx(r)>1$ at small scales.  While a correlation
coefficient defined in terms of smoothed density fields, e.g. 
$\langle \delta_g(\vx)\delta_m(\vx)\rangle /
 [\langle \delta_g^2(\vx) \rangle \langle \delta_m^2(\vx) \rangle]^{1/2}$,
must lie in the range $[-1,1]$, the same is not true for $\rx(r)$,
which is defined by ratios of correlation functions.
The high values of $\rx(r)$ at small $r$ reflect the location
of a significant fraction of galaxies (especially in the $L_*$
sample) near the central density maxima of dark matter halos.
Galaxy-mass and galaxy-galaxy pairs involving these central galaxies
boost $\xigm(r)$ and $\xigg(r)$, respectively, relative to the values
expected if galaxies traced mass within halos.  There is no
corresponding boost to $\ximm(r)$, and 
since $\rx(r)\propto \xigm(r)/[\xigg(r)]^{1/2}$, it tends
to rise above unity on scales where central galaxies make major
contributions to the correlation functions
(for related discussions, see \citealt{seljak00} and \citealt{berlind02}).
      
\cite{hoekstra01} measure a ratio of galaxy-mass and galaxy-galaxy
fluctuations that is directly analogous to $\bxix(r)$,
though it is defined in compensated apertures rather than annuli
$r \rightarrow r+\Delta r$.  In our terminology, their results imply
$\bxix = 1.26^{+0.15}_{-0.12}$ for our flat, $\Omega_m=0.4$ cosmology
(converted from their $b/r_\times=1.05$ for $\Omega_m=0.3$ using their
equation 14), with no detectable scale-dependence in the range
$r \sim 0.1 - 1.5\hmpc$.  This result is intermediate between that
of our complete and $L_*$ galaxy samples at $z=0.5$.
(The median redshift of the Hoekstra et al.\ lens sample is $z\sim 0.35$, 
where the bias should be intermediate between our $z=0$ and $z=0.5$ values.)
While we have not
modeled the Hoekstra et al.\ procedure in detail, our prediction of
a mild positive bias with no substantial scale dependence and a
bias factor similar to that applying on larger scales appears to
be fully compatible with their results.  The approximate
scale-independence of $\bxix(r)$ also agrees with Wilson et al.'s 
(\citeyear{wilson01a}) finding of similar shapes for the galaxy-galaxy
and galaxy-mass angular correlation functions.  It is difficult for us to 
compare the amplitude of $\bxix$ to Wilson et al.'s result, in part because
they weight galaxies by flux, and in part because they interpret their
measurement in terms of a value of $\Omega_m$ {\it assuming} that their
(red, luminous) lens galaxy population traces the mass.  The existence 
of positively and negatively biased samples with approximately
scale-independent $\bxix(r)$ in Figure~\ref{fig:gmxisub} demonstrates
that a constant ratio $\xigg(r)/\xigm(r)$ does not, on its own, justify
this assumption.

\begin{figure}
\centerline{
\epsfxsize=5.0truein
\epsfbox[60 500 530 720]{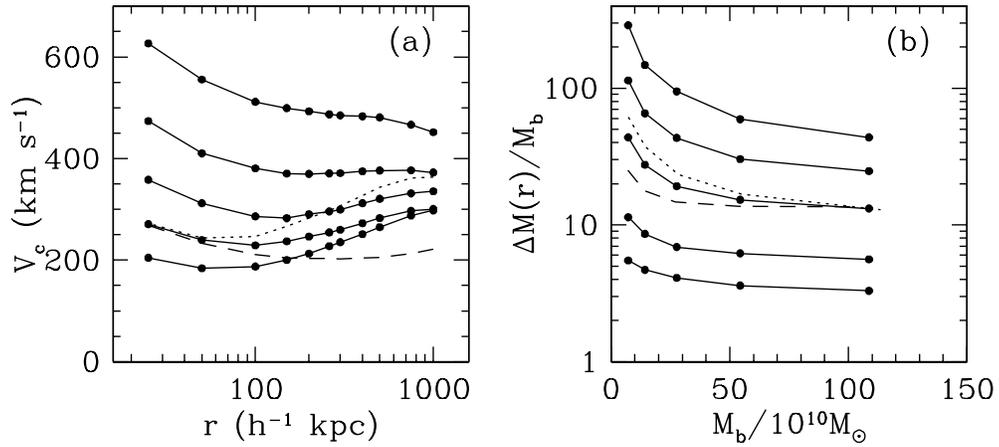}
}
\caption { 
The mass distribution around galaxies at $z=0$.
($a$) Solid curves show the circular velocity
$v_c(r) \equiv \sqrt{G\Delta M(r)/r}$ as a function of radius
where $\Delta M(r)$ is the mean total excess mass in spheres of radius $r$
centered on galaxies.  From bottom to top, the curves show
results for galaxies with baryon mass (stars plus cold gas) in the range
$0.5-1$, $1-2$, $2-4$, $4-8$, and $8-16$, in units
of $10^{11}M_\odot$.  Dotted and dashed curves show results
for, respectively, the older and younger halves of the
galaxies in the $1-2\times 10^{11}M_\odot$ mass range.
($b$) Solid curves show the ratio of the 
mean total excess mass within radius $r$ to the baryon mass of
the central galaxy as a function of the galaxy baryon mass,
for radii $r=50$, 100, 260, 500, and $1000\hkpc$ (bottom to top).
Dotted and dashed curves show results at $r=260\hkpc$ for
galaxies that are older and younger, respectively, than the median age of
galaxies in the mass bin.
}
\label{fig:massgal}
\end{figure}

\begin{figure}
\centerline{
\epsfxsize=4.0truein
\epsfbox[95 416 460 720]{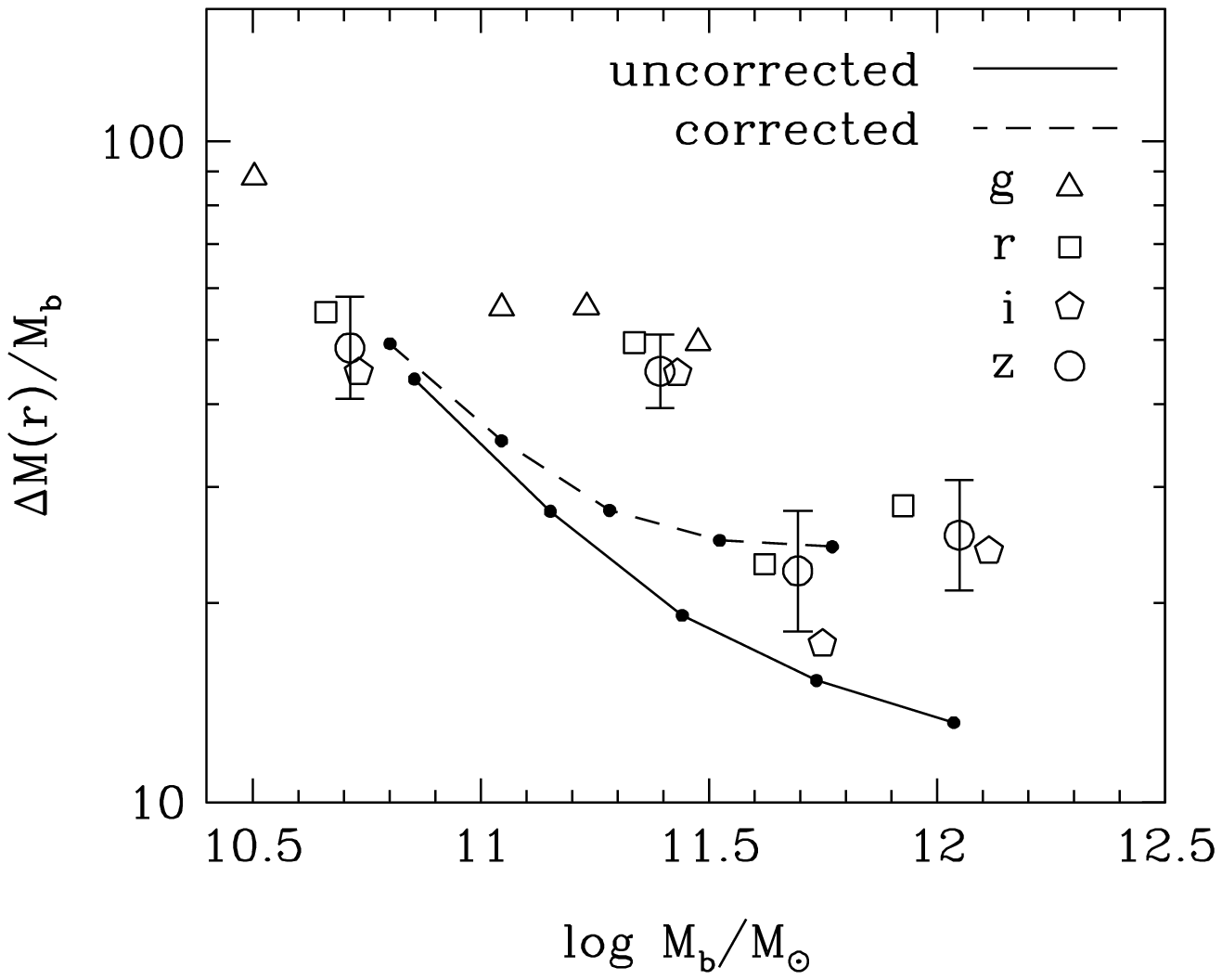}
}
\caption { 
Comparison of the simulation results for $\Delta M(260\hkpc)/M_b$
to observational estimates from the SDSS.  The solid curve (same
as the central solid curve in Fig.~\ref{fig:massgal}b) shows
the prediction from our $(50\hmpc)^3$ simulation.  The dashed 
curve shows this prediction with baryon masses rescaled to account
for numerical resolution effects, based on our comparison of 
higher resolution simulations of smaller volumes.  It represents
our best guess at the result we would obtain from a simulation
using $2\times 288^3$ particles instead of $2\times 144^3$.
Points show the weak lensing results of \cite{mckay01}, converted from 
$\Delta M(260\hkpc)/L$ to $\Delta M(260\hkpc)/M_b$ using the
estimates of $\mstel/L$ as a function of $L$ from 
\cite{kauffmann02}.  Estimates from $g$, $r$, $i$, and $z$
band are represented by triangles, squares, pentagons, and
circles, respectively.  We plot error bars on the $z$-band
points only, but they are similar for other bands.
}
\label{fig:mckay}
\end{figure}

One can also use galaxy-galaxy lensing to infer the average mass within
a sphere of radius $r$ centered on a sample galaxy, an approach
emphasized by \cite{mckay01}.  (Note that this measurement makes no attempt to 
distinguish mass that ``belongs'' to the central galaxy from mass that
``belongs'' to its neighbors or to the common halo of its group or cluster;
all mass counts, and it may be counted multiple times for different galaxies.)
Figure~\ref{fig:massgal}a shows the average mass profile from
$25\hkpc$ to $1\hmpc$ in five different bins of galaxy baryon mass,
each a factor of two in width.  We express the mass in terms of
an equivalent circular velocity $v_c = \sqrt{G\Delta M(r)/r}$, which 
would be constant in the case of a singular isothermal sphere, though
of course few galaxies would have any tracers in circular rotation
at these large radii.  We use the excess mass $\Delta M$, subtracting
the mean value $4\pi r^3 {\bar \rho}/3$, because it is the excess mass that 
is constrained by lensing, but the difference is negligible
on the scale of these plots.

For the lowest mass galaxies, $M_b= (0.5-1)\times 10^{11} M_\odot$,
the mean circular velocity curve is fairly flat at $v_c \sim 200\kms$
out to $r \sim 100\hkpc$, but it rises at larger radii because
these galaxies typically reside in groups whose characteristic
circular velocities are larger than those of the galaxies themselves.
Conversely, the highest mass galaxies have a mean circular velocity
curve that falls all the way out to $r=1\hmpc$, because these massive
galaxies typically reside at the centers of potential wells whose
mass profiles (baryon + dark) are steeper than isothermal.  Intermediate
mass ranges have intermediate behavior, with falling $v_c$ profiles
at small $r$ changing to flat or rising profiles at $r\sim 100-200\hkpc$.
Dashed and dotted curves show $v_c(r)$ for the older and younger
halves, respectively,
of the $(1-2)\times 10^{11} M_\odot$ sample.
At small radii ($r \la 50\hkpc$), the circular velocity is the same for
the two subsamples, but the preferential location of older galaxies
in more massive groups manifests itself as a higher $v_c$ at larger radii.

Figure~\ref{fig:massgal}b plots the ratio of the aperture mass
$\Delta M(r)$ to galaxy baryon mass $M_b$ (which is approximately equal
to the stellar mass), as a function of $M_b$.  From bottom to top,
solid points connected by solid lines show results for $r=50$, 100, 260,
500, and $1000\hkpc$.  
We use $260\hkpc$ rather than $250\hkpc$ to facilitate comparison
to McKay et al.'s (\citeyear{mckay01}) analysis of galaxy-galaxy
lensing in the SDSS.
We average in the order appropriate for such a study,
computing the mean value of the excess mass in bins of $M_b$, dividing by the
mean $M_b$ of the bin, and plotting the result for each bin at the
value of the mean $M_b$.  Note that this ratio counts all excess mass 
within radius $r$, including that associated with neighboring galaxies,
but divides by the baryon mass of the central galaxy only. 
For $r=50\hkpc$ the $\Delta M(r)/M_b$ 
curve is essentially flat --- on average, a galaxy's
baryon mass is proportional to the total mass in a $50\hkpc$ sphere
around it.  For larger $r$, the ratio rises towards smaller $M_b$,
which makes sense if the apertures around these galaxies start to
include more mass that ``belongs'' to their richer neighbors.
Dotted and dashed lines are computed for the older and younger halves
of the galaxies in each $M_b$ bin, at $r=260\hkpc$.  
For $M_b \la 5\times 10^{11}M_\odot$, the older galaxies have
substantially higher $\Delta M(260\hkpc)$, by a factor $\sim 1.5-2$,
analogous to the factor $\sim 2.7$ difference that \cite{mckay01}
find for elliptical and spiral subsamples.
At higher masses, there is little difference in age between the older
and younger halves of the sample (essentially all massive galaxies
are old), and the systematic difference in $\Delta M(260\hkpc)$ narrows.

Figure~\ref{fig:mckay} compares our predictions to the results of
\cite{mckay01}, who estimate $\Delta M(260\hkpc)$ by fitting isothermal
spheres to their measured shear profiles in four wide bins of galaxy
luminosity.  They find that $\Delta M(260\hkpc)/L$ is roughly 
independent of galaxy luminosity $L$ in the Sloan $g$, $r$, $i$,
and $z$ bands.  To convert $\Delta M(260\hkpc)/L$ to 
$\Delta M(260\hkpc)/M_b$ for comparison with our predictions,
we draw on the results of \cite{kauffmann02}, who estimate ratios
of stellar mass to observed luminosity by fitting population synthesis
and dust extinction models to SDSS galaxy spectra and colors.
The trend of mean stellar mass-to-light ratio with galaxy luminosity is
well described by the relations
$\langle \mstel/L\rangle = \Upsilon_* (L/L_*)^\beta$, with
$(\Upsilon_*,\beta)=$ 
(2.65,0.375) in $g$, 
(2.29,0.375) in $r$,
(2.06,0.35) in $i$, and
(1.57,0.275) in $z$, where
$L_*=1.11$, 1.51, 2.05, and $2.58 \times 10^{10} h^{-2} L_\odot$
in the four bands, respectively (\citealt{blanton01}, 
for $\Omega_m=0.3$, $\Omega_\Lambda=0.7$).\footnote{Kauffmann et al.'s
Figure~14 plots median $\mstel/L$ values and interquartile ranges
in bins of absolute magnitude.  Since it is the mean $\mstel/L$ that
matters for our purposes, we have computed these means using the
tabulated $\mstel/L$ distributions kindly provided by G.\ Kauffmann,
excluding the 10\% tails of the distributions so that the mean
is not distorted by extreme $\mstel/L$ values that may be a result
of poor model fits.  Our conclusions would not be substantially
different if we instead used the median $\mstel/L$ values plotted
by \cite{kauffmann02}.}
These relations fit the \cite{kauffmann02} results to within $\sim 10\%$
for galaxies within two magnitudes of $L_*$, though they break down
outside of this range.
We take the values of $\Delta M(260\hkpc)/L$ from McKay et al.'s Table 2,
convert to $h=0.65$, and divide by the value of $\langle \mstel/L\rangle$
at the central luminosity of each bin (which McKay et al.\ compute using
the same galaxy weights that enter the lensing analysis).
In Figure~\ref{fig:massgal}b, triangles, squares, hexagons, and circles
show results derived from the $g$, $r$, $i$, and $z$ band mass-to-light
ratios, respectively.  Results in the four bands are not independent
because they rely on the same lensing data, but the generally good
agreement among them suggests that there are no major systematic
errors in the population synthesis and dust extinction modeling.
The data points come in four groups reflecting the four luminosity
bins, though the $g$ band points do not reach the highest $M_b$
values because of the higher $\mstel/L$ in $g$ band.
The solid line in Figure~\ref{fig:mckay} shows the simulation result
for all galaxies (same as the middle solid line of Fig.~\ref{fig:massgal}b).

The predicted and observationally inferred values of $\Delta M(260\hkpc)/M_b$
agree well in the lowest mass bin, but at higher masses the observational
points lie above the simulation predictions, by a
factor $\sim 1.5-2$ on average.  
We have treated our galaxies as entirely stellar,
and we have ignored the light ``missed'' by SDSS Petrosian luminosities
(see \citealt{strauss02}), but accounting for these effects would reduce
the discrepancy by only $\sim 10-20\%$.  We conclude that the baryon
masses of our simulated galaxies are on average too high by a factor
$\sim 1.5-2$, for $M_b \ga 2\times 10^{11} M_\odot$, provided that
the \cite{kroupa02} stellar IMF assumed in Kauffmann et al.'s 
(\citeyear{kauffmann02}) population synthesis
modeling is indeed universal across galaxy
types and evolutionary stages.  Other hints of a similar discrepancy
include the high overall fraction of stellar mass in the simulation
\citep{dave01} and the relatively high stellar mass-to-light ratios
required to match the observed galaxy luminosity function
(Katz et al., in preparation).

We believe that these excessive galaxy baryon masses are largely numerical
in origin.  A comparison of two simulations of the same $(22.22\hmpc)^3$
volume, one with the same resolution as the $(50\hmpc)^3$ simulation
analyzed here and one with a mass resolution eight times higher,
shows that the {\it same} galaxies are on average more massive in the lower
resolution simulation, by $\sim 10\%$ at the $5\times 10^{10}M_\odot$
completeness limit rising to a factor $\sim 2$ at masses of 
several$\,\times\,10^{11}M_\odot$ (Fardal et al., in preparation;
one can see hints of this effect in the mass function comparison
of Figure~\ref{fig:massfun}).  The higher baryon masses at low resolution 
probably arise from the overestimated cooling rate of hot gas particles
in the vicinity of cold, dense clumps \citep{pearce99,croft01}.
An alternative formulation of SPH that performs better in the presence
of contact discontinuities between cold and hot phases indeed leads to
lower galaxy baryon masses, typically by a factor $\sim 2$
\citep{springel02}.
Empirically, we find that the relation 
$M_{\rm hr}=5\times 10^{10}M_\odot 
(M_{\rm lr}/5.4\times 10^{10}M_\odot)^{0.82}$
provides a fairly accurate scaling between masses in the ``low resolution''
and ``high resolution'' simulations.  The dashed curve in 
Figure~\ref{fig:mckay} shows the effect of applying this empirical
scaling.  It represents our best guess at the results we would
obtain from a $2\times 288^{3}$ particle simulation of the same
$(50\hmpc)^3$ volume.  Comparison to still higher resolution simulations
at $z=3$ suggests that this is sufficient to yield numerically converged
baryon masses, but we have not yet been able to carry out such a 
test at $z=0$.

Clearly, there are still systematic uncertainties in the observational
results (the weak lensing measurements themselves and the $\mstel/L$
scaling) and in the numerical convergence of the simulation
predictions.  However, the resolution-corrected predictions and
the observationally inferred mass ratios agree well at 
log$M_b/M_\odot\sim 10.7$ and 11.7, with a factor $\sim 2$ gap at
the intermediate mass scale log$M_b/M_\odot\sim 11.4$.
The sharp drop between the second and third groups of observational data points
is somewhat surprising, at least given the smooth predicted trend,
so a full assessment of the comparison should await improved
observational estimates, which should be available soon.
Nonetheless, the current level of overall agreement 
does not leave much room for additional astrophysical
processes, such as AGN feedback or stronger stellar feedback, to
substantially reduce the baryon masses of galaxies with
$M_b \ga 5\times 10^{10}M_\odot$.
We would probably predict lower values of $\Delta M(260\hkpc)/M_b$
(and thus require some suppression of baryon mass) if we adopted
lower $\Omega_m$ or higher $\Omega_b$, without changing the assumed
$\sigma_8$ or $P(k)$ shape.  However, for our adopted cosmological
parameters, the lensing comparison suggests that this simulation already
incorporates the astrophysics needed to understand the baryon
masses of luminous galaxies, within the current numerical and observational
uncertainties.  The constraints from galaxy-galaxy lensing complement
those from the Tully-Fisher relation and the luminosity function, which
we will discuss elsewhere (Katz et al., in preparation).

\section{Summary and Discussion}\label{sec: disc}

For the full population of 
galaxies with $M_b > 5\times 10^{10}M_\odot$, our simulation 
predicts a bias factor $b_\sigma(8\hmpc)$ of unity at $z=0$.
Beneath this rms similarity of fluctuation amplitudes, however,
lie numerous differences between the galaxy and dark matter
distributions, in terms of present-day structure and the evolution
of that structure.

1. The bias defined by rms fluctuations or by correlation functions
is strongly redshift dependent.  If we choose a galaxy population of
fixed comoving space density (above a redshift-dependent baryon mass
threshold), then the galaxy correlation function stays approximately
constant from $z=3$ to $z=0$, so that bias is initially large and
declines towards unity as the dark matter catches up to the galaxies
(see Figure~\ref{fig:corrall}).

2. The bias is scale-dependent in the non-linear regime, and this scale
dependence itself depends on redshift.  At $z=0$, the galaxy correlation
function is depressed below the dark matter correlation function
at $r\sim 0.5\hmpc$ and boosted above it at $r \la 0.1\hmpc$.
As a consequence, the galaxy correlation function is well described
by a power law (of slope $\gamma = 1.78 \pm 0.05$), while the
dark matter correlation function is not (see Figure~\ref{fig:powerlaw}).
The dark matter correlation function is shallower at high redshift
while the galaxy correlation function slope is roughly constant,
so the rms bias factors at $2\hmpc$ and $8\hmpc$ start to diverge
at $z\ga 1$ (see Figure~\ref{fig:bsigma}).

3. The correlation amplitude depends on galaxy baryon mass, with only
a slight difference between the more and less massive halves of the
complete sample (Figure~\ref{fig:corrsub}b) but significant increases
of $r_0$ and $b_\sigma$ result if we select the 500 or 200 most 
massive galaxies
in the simulation volume (Figures~\ref{fig:r0gamma} and~\ref{fig:bsigma},
Table~\ref{tab:subsample}).

4. The dependence of the correlation amplitude on stellar population
age is more striking than the dependence on baryon mass.  The
correlation functions of the older and younger halves of the
complete sample have almost identical slopes, but their correlation
lengths differ by nearly a factor of two (Figure~\ref{fig:corrsub},
Table~\ref{tab:subsample}).

5. At $z=0$, the pairwise velocity dispersion $\sigma_{12}(r)$ of
galaxies is lower than that of the dark matter, by $\sim 20\%$
(Figure~\ref{fig:pv}).  There is a large difference between
the pairwise dispersions of old and young galaxies, with the 
preferential location of old galaxies in dense environments
making $\sigma_{12}(r)$ about twice as high at $r \la 2\hmpc$.
The mean pairwise velocity $v_{12}(r)$ is similar for galaxies
and dark matter at $z=0$, but the denser environments of
older galaxies again cause a higher $v_{12}(r)$.  At $z=1$,
the positive spatial bias of the galaxy population makes the mean
pairwise velocity higher than that of the dark matter,
though the pairwise dispersion remains somewhat lower.

6. The skewness of galaxy count fluctuations, as quantified
by the hierarchical ratio $\stf$, is lower than that
of the dark matter by $\sim 20\%$ (Figure~\ref{fig:s3}).
The $\stf$ ratio and its scale dependence vary with galaxy
baryon mass and age, in a fairly complex way.

7. The galaxy-mass cross-correlation function depends on galaxy
mass and age in much the same way that the galaxy-galaxy correlation
function itself does, though the trends are weaker because
the mass distribution is the same in all cross-correlations
(Figure~\ref{fig:gmxisub}).  For each class of galaxies,
the bias function defined by the ratio $\xigg(r)/\xigm(r)$ is
roughly scale-independent, even when it is significantly
different from unity, though it can be somewhat different
in amplitude and shape
from the analogous bias function $[\xigg(r)/\ximm(r)]^{1/2}$
(Figure~\ref{fig:gmxibias}).

8. The average extended mass distributions around galaxies depend
significantly on baryon mass and age (Figure~\ref{fig:massgal}).
The distribution around a typical high mass galaxy falls more steeply
than an isothermal distribution, so that circular velocity curves
$[G\Delta M(r)/r]^{1/2}$ decline from $r=25\hkpc$ out
to $r=1\hmpc$.  For low mass galaxies, on the other hand, the
average mass profile is shallower than isothermal at $r \ga 100\hkpc$,
where the individual galaxy halos begin to encounter the environment
of a typical surrounding group.  This effect is much more pronounced
for old galaxies than for young galaxies.  The ratio of mean
aperture mass $\Delta M(r)$ to galaxy baryon mass $M_b$ is roughly
independent of baryon mass for apertures ranging from $r=50\hkpc$
to $r=1\hmpc$, though the ratio is higher for the galaxies with
the lowest $M_b$.  The ratio is higher for old galaxies than for
young galaxies, again demonstrating the preferential location of
old galaxies in dense environments.

For the most part, the dark matter clustering of the $\Lambda$CDM
model and these biases of the galaxy population lead to good qualitative
and, where we have adequate statistics for comparison, quantitative
agreement with observed galaxy clustering.  The predicted galaxy 
correlation function is a power law of the observed slope and approximately
the observed amplitude (see further discussion below).
The redshift-independence of the comoving correlation length agrees
with measurements from Lyman-break galaxy surveys at $z\approx 3$
\citep{adelberger98,adelberger02} and from deep redshift surveys
extending to $z\sim 1$ \citep{postman98,carlberg99}.
The predicted dependence of clustering strength on baryon mass
and population age agrees with the trends as a function
of luminosity, color, and spectral type found in
the 2dFGRS and SDSS \citep{norberg01,norberg02b,zehavi02}.
The predicted value of the hierarchical ratio $\stf$ agrees with
measurements from angular clustering catalogs 
\citep{gaztanaga94,gaztanaga02,szapudi02}.  The value and
scale-independence of the bias inferred from the ratio of
galaxy-galaxy and galaxy-mass correlation functions agrees
with recent observational estimates \citep{hoekstra01}.
The dependence of the galaxy-mass correlation function on population
age parallels the dependence on galaxy type found by \cite{mckay01},
and the aperture mass $\Delta M(260\hkpc)$ is approximately proportional
to galaxy baryon mass, as \cite{mckay01} also find.

There are three quantitative
discrepancies between our predictions and existing measurements,
two fairly subtle and one more substantial.
First, the power law slope of $\xi(r)$ for the $L_*$-galaxy sample
(the 500 most massive galaxies in the box at $z=0$) is slightly
too steep, $\gamma=2.00\pm 0.08$ compared to observed values of $1.7-1.8$,
and there are hints of a departure from a power law at $r \la 1\hmpc$.
Second, the correlation length of this sample, $r_0=4.5\pm 0.4\hmpc$,
is low compared to the value $r_0=6.3\pm 0.8\hmpc$ measured for
$r$-selected galaxies with $L \approx L_*$ in the SDSS \citep{zehavi02},
though it is consistent with the value $r_0=4.9\pm 0.3\hmpc$ found
for $L_*$ $b_{\rm J}$-selected galaxies in the 2dFGRS \citep{norberg01}.
Because of the still-limited size of our simulation volume, these
discrepancies are of marginal statistical significance, and 
the N-body tests discussed in \S\ref{sec:corrall} imply that the
low correlation length, at least, is a consequence of the lower
than average fluctuation amplitude in the particular realization
of $\Lambda$CDM initial conditions used here.

The more serious quantitative discrepancy is the ratio of aperture
mass $\Delta M(260\hkpc)$ to galaxy baryon mass $M_b$.
For $M_b \sim (0.5-1) \times 10^{11} M_\odot$, the simulation prediction
agrees with the result derived by combining McKay et al.'s (\citeyear{mckay01})
values of $\Delta M/L$ with Kauffmann et al.'s (\citeyear{kauffmann02})
stellar mass-to-light ratios, but at higher $M_b$ the predicted ratios
are lower than the observationally inferred values by factors of $\sim 1.5-2$,
implying that the galaxy baryon masses in the simulation are too high
(Figure~\ref{fig:mckay}).
As discussed in \S\ref{sec: gdm}, we believe that this discrepancy
is mostly numerical in origin, since our internal resolution tests
(Fardal et al., in preparation) and comparisons between different
SPH implementations \citep{springel02} suggest that galaxy baryon masses
are indeed overestimated by about this factor in this mass range,
relative to a simulation with higher resolution or more accurate treatment
of the interface between hot and cold gas phases.
A comparison between our best guess at resolution-corrected results
and the \cite{mckay01} data points yields fairly good agreement
in the average values of $\Delta M(260\hkpc)/M_b$, which leaves
little room for additional astrophysical processes,
such as more aggressive feedback, to substantially suppress gas
cooling and star formation in this galaxy mass range.
Consistent with this conclusion, \cite{yang02} find that the 
N-body + semi-analytic models of \cite{kauffmann99a}, which do include
much stronger stellar feedback, predict $\Delta M(260\hkpc)/L$
ratios that are too low (by a factor $\sim 2$) relative to those
of \cite{mckay01}.  As emphasized by
\cite{guzik02} and \cite{seljak02a,seljak02b}, the constraints
from galaxy-galaxy lensing measurements complement those from the
galaxy luminosity function and the Tully-Fisher and fundamental
plane relations because they suffer different systematic uncertainties
and have different sensitivity to astrophysical and cosmological parameters.
For example, lowering $\Omega_m$ while keeping other parameters
(including the linear power spectrum) fixed would tend to lower
$\Delta M(r)/L$ while having relatively little impact on the
luminosity function (though galaxy formation, and hence galaxy 
luminosities, would still be affected by the different timescales
and baryon-to-dark matter ratios of the lower $\Omega_m$ model).
The weak lensing constraints will become considerably more powerful as
the precision and detail of the measurements improves.

Where we have been able to compare, our results generally agree well
with those of other hydrodynamic simulations 
\citep{pearce99,pearce01,cen00,yoshikawa01}, high resolution
N-body simulations that identify galaxies with sub-halos
\citep{colin99,kravtsov99}, and combined N-body + semi-analytic
models \citep{kauffmann97,governato98,kauffmann99a,kauffmann99b,
benson00a,benson00b,benson01}.
All of these methods make qualitatively similar predictions
for the evolution of the galaxy correlation function,
the bias between galaxies and mass at the present day,
the dependence of that bias on galaxy type,
and the relative amplitude of dark matter and galaxy pairwise dispersions.
This agreement suggests that these aspects of galaxy clustering
and bias arise from robust aspects of the physics of galaxy
formation that all of these methods treat correctly, or at least 
similarly.  Our conclusions about $\stf$ and galaxy-mass correlations are
for the most part new, but we do not expect them to be fundamentally
different from the predictions of other methods.

Returning to the main goals of the paper, we find that the combination
of a $\Lambda$CDM cosmology with standard ideas about galaxy formation
is, on the whole, remarkably successful at reproducing observed
galaxy clustering and galaxy-galaxy lensing measurements.
In terms of physical interpretation, the simulation analysis offers
a number of insights.
It shows that the detailed relation between the galaxy and dark halo
populations can account for the difference between the predicted 
correlation function of dark matter and the observed power law form
of the galaxy correlation function, and between the predicted pairwise
velocity dispersion of the dark matter and the observed, lower dispersion
of galaxies.
It shows that the observed dependence of galaxy clustering on galaxy
properties emerges naturally from the dependence of galaxy mass and
population age on environment --- basically, older and more massive
galaxies form in regions that collapse early and are today biased
with respect to the overall mass distribution.  It shows that values
of $b_\sigma \sim 1$ and $\bxix \sim 1$ can emerge even when galaxies
do not trace mass in detail.  The quantitative
discrepancies discussed above may
also prove instructive, as the observations improve and the uncertainties
in the simulations themselves are more thoroughly understood.

Our analysis makes a number of predictions that can be tested with future
data, or with future analyses of existing data.
Our quantitative comparison to the observed type dependence of galaxy
clustering has been limited by our finite simulation volume, which
prevents us from predicting the clustering of rare classes of galaxies.
However, it would be straightforward to test the quantitative predictions
presented in, for example, 
Figure~\ref{fig:corrsub} by taking an observed sample 
above a luminosity threshold that yields the same space density as our 
complete sample and dividing it in two based on luminosity or color
(or, better still, on the basis of stellar mass and population age,
using techniques like those of \citealt{kauffmann02}).
The predicted evolution of the mean pairwise velocity and pairwise dispersion
(Figure~\ref{fig:pv})
can be tested with upcoming large redshift surveys like the DEEP and 
VIRMOS-VLT programs.
The systematic variations of $\stf$ with galaxy luminosity and age
(Figure~\ref{fig:s3}) can be tested with the 2dFGRS and SDSS.
The most informative comparisons will probably come from improved
measurements of galaxy-galaxy lensing and cosmic shear, which will
allow detailed tests of the trends predicted in Figures~\ref{fig:gmxibias}
and~\ref{fig:massgal}.

On all of these fronts, observational efforts are advancing at
a staggering pace.  Comparisons between more comprehensive data and
improved theoretical predictions over the next few years should
tell us whether our understanding of dark matter clustering
and the physics of galaxy formation is indeed complete, at 
least in terms of features that have major quantitative impact,
or whether there are still important ingredients missing from
our theoretical recipe.

\acknowledgments
We thank Andreas Berlind, Stephane Colombi, Mark Fardal, Henk Hoekstra, 
Guinevere Kauffmann, Tim McKay, Yannick Mellier, and Ludo van Waerbeke 
for valuable discussions on numerous aspects of this study.
This work was supported by the NSF and by the NASA Astrophysical
Theory Program and Long-Term Space Astrophysics Program.
RD acknowledges the support of Hubble Fellowship grant number 
HST-HF-01128.01-A
from the Space Telescope Science Institute, which is operated by
AURA, Inc., under NASA contract NAS5-26555.
DHW acknowledges the hospitality
of the Institute for Advanced Study and the Institut d'Astrophysique de Paris,
and the support of the Ambrose Monell Foundation and the 
French C.N.R.S., during the completion of this work.



%
%


\begin{thebibliography}{}


\bibitem[Adelberger et al.(1998)]{adelberger98} 
Adelberger, K. L. , Steidel, C. C., Giavalisco, M., 
Dickinson, M., Pettini, M., \& Kellogg, M. 1998, \apj, 505, 18

\bibitem[Adelberger et al.(2002)]{adelberger02} 
Adelberger, K. L. , Steidel, C. C., Shapley, A., \& Pettini, M. 2002, \apj, 
in press, astro-ph/0210314


\bibitem[Bahcall et al.(1999)]{bahcall99} 
Bahcall, N., Ostriker, J. P., Perlmutter, S., \& Steinhardt, P. J. 1999, 
Science, 284, 1481

\bibitem[Bean et al.(1983)]{bean83}
Bean, A. J., Ellis, R. S., Shanks, T., Efstathiou, G., \& Peterson, B. A. 1983,
\mnras, 205, 605

\bibitem[Bennett et al.(1996)]{bennett96} 
Bennett, C. L., Banday, A. J.,
Gorski, K. M., Hinshaw, G., Jackson, P., Keegstra, P., Kogut, A., Smoot, G. F.,
Wilkinson, D. T., \& Wright, E. L. 1996, \apjl, 464, L1

\bibitem[Benoist et al.(1996)]{benoist96} 
Benoist, C., Maurogordato, 
da Costa, L. N., Cappi, A., \& Schaeffer, R. 1996, \apj, 472, 452

\bibitem[Benson et al.(2000a)]{benson00a}
Benson, A.~J., Baugh, C.~M., Cole, S., Frenk, C.~S., \&
Lacey, C.~G.\ 2000a, \mnras, 316, 107

\bibitem[Benson et al.(2000b)]{benson00b}
Benson, A. J., Cole, S., Frenk, C. S., Baugh, C. M., \& Lacey, C. G. 2000b,
\mnras, 311, 793

\bibitem[Benson et al.(2001)]{benson01}
Benson, A.~J., Frenk, C.~S., Baugh, C.~M., Cole, S., \& Lacey, C.~G.\ 2001,
\mnras, 327, 1041

\bibitem[Berlind et al.(2002)]{berlind02b}
Berlind, A. A., Weinberg, D. H., Benson, A. J., Baugh, C. M.,
Cole, S., Dav\'e, R., Frenk, C. S., Jenkins, A., Katz, N., \&
Lacey, C. G. 2002, \apj, submitted

\bibitem[Berlind \& Weinberg(2002)]{berlind02}
Berlind, A. A., \& Weinberg, D. H. 2002, \apj, 575, 587

\bibitem[Bertschinger (1998)]{bertschinger98} 
Bertschinger, E. 1998, ARA\&A, 36, 599

\bibitem[Blanton et al.(2001)]{blanton01} 
Blanton, M.R. et al. 2001, \aj, 121, 2358

\bibitem[Blumenthal et al.(1984)]{blumenthal84} 
Blumenthal, G. R., Faber, S. M., 
Primack, J. R., \& Rees, M. J. 1984, Nature, 311, 517

\bibitem[Bond et al.(1991)]{bond91} 
Bond, J. R., Cole, S., Efstathiou, G., \&
Kaiser, N. 1991, \apj, 379, 440

\bibitem[Bouchet et al.(1993)]{bouchet93}
Bouchet, F. R., Strauss, M. A., Davis, M., Fisher, K. B.,
Yahil, A., \& Huchra, J. P. 1993, \apj, 417, 36

\bibitem[Bower (1991)]{bower91} 
Bower, R. J. 1991, \mnras, 248,, 332

\bibitem[Brainerd, Blandford, \& Smail(1996)]{brainerd96}
Brainerd, T. G., Blandford, R. D., \& Smail, I. 1996, \apj, 466, 623

\bibitem[Burles \& Tytler(1998)]{burles98}
Burles, S., \& Tytler, D. 1998, \apj, 499, 699

\bibitem[Carlberg \& Couchman(1989)]{carlberg89}
Carlberg, R.~G.~\& Couchman, H.~M.~P.\ 1989, \apj, 340, 47.

\bibitem[Carlberg et al.(1999)]{carlberg99} 
Carlberg, R., et al.  1999,
Phil. Trans. R. Soc. Lond. A, 356, 167

\bibitem[Cen \& Ostriker(1992)]{cen92}
Cen, R., \& Ostriker, J. P. 1992, \apj, 399, L113

\bibitem[Cen \& Ostriker(2000)]{cen00} 
Cen, R. \& Ostriker, J. P. 2000, \apj, 538, 83

\bibitem[Chen et al.(2002)]{chen02}
Chen, X., Weinberg, D.~H., Katz, N., \& Dav\'e, R.\ 2002, \apj, submitted,
astro-ph/0203319

\bibitem[Col\'{\i}n et al.(1999)]{colin99} 
Col\'in, P., Klypin, A. A., 
Kravtsov, A. V., \& Khokhlov, A. M. 1999, \apj, 523, 32

\bibitem[Colombi et al.(2000)]{colombi00}
Colombi, S., Szapudi, I., Jenkins, A., \& 
Colberg, J.\ 2000, \mnras, 313, 711

\bibitem[Croft et al.(2001)]{croft01}
Croft, R. A. C., Di Matteo, T., Dav\'e, R., Hernquist, L., Katz, N.,
Fardal, M., \& Weinberg, D. H. 2001, \apj, 557, 67

\bibitem[Croft et al.(1999)]{croft99}
Croft, R. A. C., Weinberg, D. H., Pettini, M., Katz, N., \& Hernquist, L. 1999,
\apj, 520, 1

\bibitem[Croft et al.(2002)]{croft02} 
Croft, R. A. C., Weinberg, D. H., Bolte, M., Burles, S.,  Hernquist, L., Katz,
N., Kirkman, D., Tytler, D. 2002, \apj, in press, astro-ph/0012324


\bibitem[Dav\'e et al.(2001)]{dave01}
Dav\'e, R., Cen, R., Ostriker, J. P., Bryan, G. L., Hernquist, L., Katz, N.,
Weinberg, D. H., Norman, M. L., \& O'Shea, B. 2000, \apj, 552, 473

\bibitem[Dav\'e, Dubinski, \& Hernquist(1997)]{dave97} 
Dav\'e, R., Dubinski, J., \& Hernquist, L. 1997a, NewAst, 2, 277

\bibitem[Dav\'e, Katz, \& Weinberg(2002)]{dave02} 
Dav\'e, R., Katz, N., \& Weinberg, D. H. 2002, \apj, 579, 23

\bibitem[Davis et al.(1985)]{davis85}
Davis, M., Efstathiou, G., Frenk, C. S., \& White,
S. D. M. 1985, \apj, 292, 371

\bibitem[Davis \& Geller(1976)]{davis76} 
Davis, M. \& Geller, M. J. 1976, \apj, 208, 13

\bibitem[Davis, Groth, \& Peebles(1977)]{davis77}
Davis, M., Groth, E.~J., \& Peebles, P.~J.~E.\ 1977, \apjl, 212, L107.

\bibitem[Davis \& Peebles(1983)]{davis83} Davis, M. \& Peebles, P. J. E. 1983,
\apj, 267, 465

\bibitem[de Lapparent, Geller, \& Huchra(1988)]{delapparent88}
de Lapparent, V., Geller, M.J., \& Huchra, J.P. 1988, \apj, 332, 44

\bibitem[dell'Antonio \& Tyson(1996)]{dellantonio96}
dell'Antonio, I. P. \& Tyson, J. A. 1996, \apjl, 473, L17

\bibitem[Burles \& Tytler(1997)]{burles97}
Burles, S., \& Tytler, D. 1997, \aj, 114, 1330


\bibitem[Diaferio et al.(1999)]{diaferio99} 
Diaferio, A., Kauffmann, G., 
Colberg, J. M., \& White, S. D. M. 1999, \mnras, 307, 537

\bibitem[Dressler(1980)]{dressler80}
Dressler, A. 1980, \apj, 236, 351

\bibitem[Efstathiou et al.(1990)]{efstathiou90}
Efstathiou, G.,
Kaiser, N., Saunders, W., Lawrence, A., Rowan-Robinson, M., Ellis, R.~S.,
\& Frenk, C.~S.\ 1990, \mnras, 247, 10P

\bibitem[Evrard, Summers, \& Davis(1994)]{evrard94}
Evrard, A.E., Summers, F.J., \& Davis, M. 1994, \apj, 422, 11

\bibitem[Eke, Cole, \& Frenk(1996)]{eke96} 
Eke, V., Cole, S., \& Frenk, C. S.
1996, \mnras, 282, 263

\bibitem[Falco et al.(1999)]{falco99} 
Falco, E. E., Kurtz, M. J.,
Geller, M. J., Huchra, J. P., Peters, J., Berlind, P.,
Mink, D. J., Tokarz, S. P., Elwell, B. 1999, \pasp, 111, 438

\bibitem[Fardal et al.(2001)]{fardal01}
Fardal, M.\ A., Katz, N., Gardner, J.\ P., Hernquist, L.,
Weinberg, D.\ H.\ \& Dav{\'e}, R.\ 2001, \apj, 562, 605

\bibitem[Fardal et al.(2002)]{fardal02}
Fardal, M.\ A., Katz, N., Weinberg, D. H., Dav\'e, R., \& Hernquist, L. 2002,
\apj, in press, astro-ph/0107290

\bibitem[Ferreira et al.(1999)]{ferreira99}
Ferreira, P.~G., Juszkiewicz, R., Feldman, H.~A., Davis, M., \& 
Jaffe, A.~H.\ 1999, \apjl, 515, L1

\bibitem[Fischer et al.(2000)]{fischer00}
Fischer, P., et al.\ 2000, \aj, 120, 1198

\bibitem[Fisher(1995)]{fisher95}
Fisher, K. B. 1995, \apj, 448, 494

\bibitem[{Fisher et al.(1994)}]{fisher94}
Fisher, K.B., Davis, M., Strauss, M.A., Yahil, A., \& Huchra, J.P.
1994, \mnras, 267, 927

\bibitem[Frenk et al.(1988)]{frenk88} 
Frenk, C. S., White, S. D. M., Davis, M., \& Efstathiou, G. 1988, \apj, 327, 
507

\bibitem[Fry(1996)]{fry96}
Fry, J. N. 1996, \apj, 461, L65

\bibitem[Fry \& Gazta\~naga(1993)]{fry93}
Fry, J. N., \& Gazta\~naga, E. 1993, \apj, 413, 447

\bibitem[Fry \& Melott(1985)]{fry85}
Fry, J. N., \& Melott, A. L. 1985, \apj, 292, 395

\bibitem[Gardner et al.(1997)]{gardner97} 
Gardner, J. P., Katz, N., Hernquist, L., \& Weinberg, D. H.  1997, \apj, 484, 
31

\bibitem[Gazta\~naga(2002)]{gaztanaga02}
Gazta\~naga, E. 2002, \apj, 580, 144

\bibitem[Gazta\~naga \& Frieman (1994)]{gaztanaga94} 
Gazta\~naga, E. \& Frieman, J. A. 1994, \apjl, 437, L13

\bibitem[Ghigna et al.(1998)]{ghigna98} 
Ghigna, S., Moore, B., Governato, F.,
Lake, G., Quinn, T., \& Stadel, J. 1998, \mnras, 300, 146

\bibitem[Giovanelli, Haynes \& Chincarini (1986)]{giovanelli86} 
Giovanelli, R., Haynes, M. P., \& Chincarini, G. L. 1986, \apj, 300, 77

\bibitem[Gott et al.(1989)]{gott89}
Gott, J. R. et al.\ 1989, \apj, 340, 625

\bibitem[Gott \& Rees(1975)]{gott75}
Gott, J.~R.~\& Rees, M.~J.\ 1975, \aap, 45, 365

\bibitem[Gott \& Turner(1977)]{gott77}
Gott, J.~R.~I.~\& Turner, E.~L.\ 1977, \apj, 216, 357

\bibitem[Gott \& Turner(1979)]{gott79}
Gott, J.~R.~\& Turner, E.~L.\ 1979, \apjl, 232, L79

\bibitem[Gott, Turner, \& Aarseth(1979)]{gott79b}
Gott, J. R., Turner, E. L., \& Aarseth, S. J. 1979, \apj, 234, 13

\bibitem[Governato et al.(1998)]{governato98}
Governato, F., Baugh, C.M., Frenk, C.S., Cole, S., Lacey, C.G., Quinn, T.R. \&
Stadel, J. 1998, Nature, 392, 359

\bibitem[Griffiths et al.(1996)]{griffiths96}
Griffiths, R. E., Casertano, S., Im, M., \& Ratnatunga, K. U. 1996,
\mnras, 282, 1159

\bibitem[Guzik \& Seljak(2001)]{guzik01}
Guzik, J., \& Seljak, U. 2001, \mnras, 321, 439

\bibitem[Guzik \& Seljak(2002)]{guzik02}
Guzik, J., \& Seljak, U. 2002, \mnras, 335, 311

\bibitem[Guzzo et al.(1997)]{guzzo97} 
Guzzo, L., Strauss, M.\
A., Fisher, K.\ B., Giovanelli, R., \& Haynes, M.\ P.\ 1997, \apj, 489, 37

\bibitem[Guzzo et al.(2000)]{guzzo00} Guzzo, L. \etal 2000, A\&A, 355, 1

\bibitem[Hamilton(1988)]{hamilton88}
Hamilton, A. J. S. 1988, \apj, 331, L59

\bibitem[Hamilton(1992)]{hamilton92}
Hamilton, A. J. S. 1992, \apj, 385, L5

\bibitem[Hamilton et al.(1991)]{hamilton91}
  Hamilton, A. J. S., Matthews, A., Kumar, P., \& Lu, E. 1991,
  \apj, 374, L1

\bibitem[Hatton et al.(2002)]{hatton02}
Hatton, S., Devriendt, J. E. G., Ninin, S., Bouchet, F. R., 
Guiderdoni, B., \& Vibert, D.\ 2002, \mnras, submitted


\bibitem[Hoekstra, Yee, \& Gladders(2001)]{hoekstra01}
Hoekstra, H., Yee, H.~K.~C., \& Gladders, M.~D.\ 2001, \apjl, 558, L11

\bibitem[Hoekstra, Yee, \& Gladders(2002)]{hoekstra02}
Hoekstra, H., Yee, H.~K.~C., \& Gladders, M.~D.\ 2002, \apj, 577, 595

\bibitem[Hudson et al.(1998)]{hudson98}
Hudson, M. J., Gwyn, S. D. J., Dahle, H., \& Kaiser, N. 1998, \apj, 503, 531

\bibitem[Jaffe et al.(2001)]{jaffe00} 
Jaffe, A., et al. 2001, Phys Rev Lett, 86, 3475

\bibitem[Jenkins et al.(1998)]{jenkins98} 
Jenkins, A., Frenk, C. S., Pearce, F. R., Thomas, P. A., Colberg, J. M., 
White, S. D. M., Couchman, H. M. P., Peacock, J. A., Efstathiou, G., \& 
Nelson, A. H. (The Virgo Consortium) 1998, \apj, 499, 20

\bibitem[Jing, Mo \& B\"orner(1998)]{jing98} 
Jing, Y. P., Mo, H. J., \& B\"orner, G.  1998, \apj, 494, 1

\bibitem[Jing, B\"orner, \& Suto(2002)]{jing02} 
Jing, Y. P., B\"orner, G., \& Suto, Y. 2001, \apj, 564, 15

\bibitem[Juszkiewicz, Bouchet, \& Colombi(1993)]{juszkiewicz93}
Juszkiewicz, R., Bouchet, F. R., \& Colombi, S. 1993, \apj, 412, L9

\bibitem[Juszkiewicz et al.(1995)]{juszkiewicz95}
Juszkiewicz, R., Weinberg, D. H., Amsterdamski, P., Chodorowski, M., \&
Bouchet, F. R. 1995, \apj, 442, 39

\bibitem[Kaiser(1984)]{kaiser84} Kaiser, N. 1984, \apjl, 284, L9

\bibitem[Kaiser(1987)]{kaiser87}
Kaiser, N. 1987, \mnras, 227, 1

\bibitem[Katz, Hernquist, \& Weinberg(1992)]{katz92} 
Katz, N., Hernquist, L., \& Weinberg, D. H. 1992, \apjl, 399, L109

\bibitem[Katz, Weinberg \& Hernquist(1996)]{katz96} 
Katz, N., Weinberg D. H., \& Hernquist, L. 1996, \apjs, 105, 19 (KWH)

\bibitem[Katz, Hernquist, \& Weinberg(1999)]{katz99} 
Katz, N., Hernquist, L., \& Weinberg D. H. 1999, \apj, 523, 463

\bibitem[Kauffman et al.(2002)]{kauffmann02}
Kauffmann, G., et al., \mnras, in press, astro-ph/0204055

\bibitem[Kauffmann \& White (1993)]{kauffmann93} 
Kauffmann, G. \& White, S. D. M.  1993, \mnras, 261, 921

\bibitem[Kauffmann, Nusser, \& Steinmetz(1997)]{kauffmann97} 
Kauffmann, G., Nusser, A., \& Steinmetz, M. 1997, \mnras, 286, 795

\bibitem[Kauffman et al.(1999a)]{kauffmann99a}
Kauffmann, G., Colberg, J. M., Diaferio, A., \& White, S. D. M. 1999a,
\mnras, 303, 188

\bibitem[Kauffman et al.(1999b)]{kauffmann99b}
Kauffmann, G., Colberg, J. M., Diaferio, A., \& White, S. D. M. 1999b,
\mnras, 307, 529

\bibitem[Kennicutt (1998)]{kennicutt98} 
Kennicutt, R. C. 1998, \apj, 498, 541

\bibitem[Kollmeier et al.(2002)]{kollmeier02}
Kollmeier, J. A., Weinberg, D. H., Dav\'e, R., \& Katz, N. 2002, 
\apj, submitted, astro-ph/0209563

\bibitem[Kravtsov \& Klypin(1999)]{kravtsov99} 
Kravtsov, A. V. \& Klypin A.  1999, \apj, 520, 437

\bibitem[Kroupa(2002)]{kroupa02}
Kroupa, P.\ 2002, Science, 295, 82

\bibitem[Loveday et al.(1995)]{loveday95} Loveday, J., Maddox, S. J., 
Efstathiou, G., \& Peterson, B. A. 1995, \apj, 442, 457

\bibitem[Ma \& Fry(2000)]{ma00}
Ma, C., \& Fry, J. N. 2000, \apj, 543, 503

\bibitem[Maddox et al.(1990)]{maddox90}
Maddox, S. J., Efstathiou, G., Sutherland, W. J., \& Loveday, J. 1990,
\mnras, 242, L43

\bibitem[Mann, Peacock, \& Heavens(1998)]{mann98}
Mann, R. G., Peacock, J. A., \& Heavens, A. F. 1998, \mnras, 293, 209

\bibitem[Marzke et al.(1995)]{marzke95}
Marzke, R.\ O., Geller, M.\ J., da Costa, L.\ N., \& Huchra, J.\ P. 1995,
\aj, 110, 477

\bibitem[McDonald et al.(2000)]{mcdonald00} 
McDonald, P., Miralda-Escud\'e, J.,
Rauch, M., Sargent, W. L. W., Barlow, T. A., Cen, R. \& Ostriker, J. P.
2000, \apj, 543, 1

\bibitem[McKay et al.(2001)]{mckay01}
McKay, T. A., et al. 2001, unpublished, astro-ph/0108013

\bibitem[Mellier et al.(2001)]{mellier01}
Mellier, Y., van
Waerbeke, L., Maoli, R., Schneider, P., Jain, B., Bernardeau, F., Erben,
T., \& Fort, B.\ 2001, in Deep Fields, Proc. of ESO/ECF/STScI
Workshop, ed. ~S. Cristiani,
A. Renzini, R.~E.~Williams, (Springer), p.~252

\bibitem[Mo, Jing, \& B\"orner(1993)]{mo93} 
Mo, H. J., Jing, Y. P., \& B\"orner, G. 1993, \mnras, 264, 825

\bibitem[Murali et al.(2002)]{murali02}
Murali, C., Katz, N., Hernquist, L., Weinberg, D. H., \& Dav\'e, R. 2002,
\apj, 571, 1

\bibitem[Narayanan, Berlind, \& Weinberg(2000)]{narayanan00}
Narayanan, V. K., Berlind, A. A., \& Weinberg, D. H. 2000, \apj, 528, 1

\bibitem[Norberg et al.(2001)]{norberg01} Norberg, P., et al. 2001, \mnras, 
  328, 64

\bibitem[Norberg et al.(2002a)]{norberg02a}
  Norberg, P., et al.\ 2002a, \mnras, 336, 907

\bibitem[Norberg et al.(2002b)]{norberg02b}
Norberg, P., et al.\ 2002b, \mnras, 332, 827

\bibitem[Park(1990)]{park90}
Park, C. 1990, \mnras, 242, L59 

\bibitem[Park et al.(1994)]{park94}
Park, C., Vogeley, M. S., Geller, M. J., \& Huchra, J. P. 1994,
\apj, 431, 569

\bibitem[Peacock \& Dodds(1996)]{peacock96}
Peacock, J. A., \& Dodds, S. J. 1996, \mnras, 280, L19

\bibitem[Peacock \& Smith(2000)]{peacock00}
Peacock, J. A., \& Smith, R. E. 2000, \mnras, 318, 1144

\bibitem[Pearce et al.(1999)]{pearce99}
Pearce, F. R., Jenkins, A., Frenk, C. S., Colberg, J. M.,
White, S. D. M., Thomas, P. A., Couchman, H. M. P., Peacock, J. A.,
\& Efstathiou, G. 1999, \apj, 521, L99

\bibitem[Pearce et al.(2001)]{pearce01} 
Pearce, F. R., Jenkins, A., Frenk, C. S., 
White, S. D. M., Thomas, P. A., Couchman, H. M. P., Peacock, J. A., \&
Efstathiou, G. 2001, \mnras, 326, 649

\bibitem[Peebles(1974)]{peebles74}
Peebles, P.\ J.\ E.\ 1974, A\&A, 32, 197

\bibitem[Peebles(1980)]{peebles80}
Peebles, P. J. E. 1980, The Large Scale Structure of the
Universe (Princeton: Princeton University Press)

\bibitem[Perlmutter et al.(1999)]{perlmutter99} 
Perlmutter, S. et al. (The SCP Collaboration) 1999, \apj, 517, 565

\bibitem[Postman \& Geller(1984)]{postman84}
Postman, M., \& Geller, M. J. 1984, \apj, 281, 95

\bibitem[Postman et al.(1998)]{postman98} 
Postman, M., Lauer, T. R., Szapudi, I., \& Oegerle, W. 1998, \apj, 506, 33

\bibitem[Quinn et al.(1997)]{quinn97}
Quinn, T., Katz, N., Stadel, J. \& Lake, G. 1997, unpublished, astro-ph/9710043

\bibitem[Ratcliffe et al.(1998)]{ratcliffe98} 
Ratcliffe, A., Shanks, T., Parker, Q. A., \& Fong, R. 1998, \mnras, 296, 173

\bibitem[Riess et al.(1998)]{riess98} 
Riess, A. et al. 1998, \aj, 116, 1009


\bibitem[Schechter(1976)]{schechter76}
Schechter, P. 1976, \apj, 203, 297

\bibitem[Scoccimarro et al.(2001)]{scoccimarro01}
Scoccimarro, R., Sheth, R. K., Hui, L., \& Jain, B. 2001, \apj, 546, 20

\bibitem[Seljak(2000)]{seljak00}
Seljak, U. 2000, \mnras, 318, 203

\bibitem[Seljak(2002a)]{seljak02a}
Seljak, U. 2002a, \mnras, 334, 797

\bibitem[Seljak(2002b)]{seljak02b}
Seljak, U. 2002b, \mnras, submitted, astro-ph/0203117

\bibitem[Sheth et al.(2001)]{sheth01}
Sheth, R. K., Hui, L., Diaferio, A., \& Scoccimarro, R. 2001, \mnras,
  326, 463

\bibitem[Smith et al.(2001)]{smith01}
Smith, D.~R., Bernstein, G.~M., Fischer, P., \& Jarvis, M.\ 2001, \apj, 
551, 643

\bibitem[Somerville \& Primack(1999)]{somerville99} 
Somerville, R. S. \& Primack, J. R. 1999, \mnras, 310, 1087

\bibitem[Somerville et al.(2001)]{somerville01}
Somerville, R.~S., Lemson, G., Sigad, Y., Dekel, A., Kauffmann, G.,
\& White, S.~D.~M.\ 2001, \mnras, 320, 289

\bibitem[Springel \& Hernquist(2002)]{springel02}
Springel, V., \& Hernquist, L.\ 2002, \mnras, 333, 649

\bibitem[Strauss et al.(2002)]{strauss02}
Strauss, M.A., et al.\ 2002, \aj, 124, 1810

\bibitem[Summers, Davis, \& Evrard (1995)]{summers95} 
Summers, F. J., Davis, M., \& Evrard, A. E. 1995, \apj, 454, 1

\bibitem[Szapudi \& Szalay(1993)]{szapudi93}
Szapudi, I.~\& Szalay, A.~S.\ 1993, \apj, 408, 43

\bibitem[Szapudi et al.(2002)]{szapudi02}
Szapudi, I. et al. 2002, \apj, 570, 75

\bibitem[Tegmark \& Peebles(1998)]{tegmark98}
Tegmark, M., \& Peebles, P. J. E. 1998, \apj, 500, L79

\bibitem[Tegmark \& Bromley(1999)]{tegmark99}
Tegmark, M., \& Bromley, B. C. 1999, \apj, 518, L69

\bibitem[Totsuji \& Kihara(1969)]{totsuji69}
Totsuji, H.~\& Kihara, T.\ 1969, \pasj, 21, 221

\bibitem[Tucker et al.(1997)]{tucker97} 
Tucker, D. L., Oemler, A., Kirshner, R. P.,
Lin, H., Schectman, S. A., Landy, S. D., Schecter, P. L., Muller, V.,
Gottlober, S., \& Einasto, J. 1997, \mnras, 285, L5

\bibitem[Tully \& Fisher(1977)]{tully77}
  Tully, R.\ B.\ \& Fisher, J.\ R.\ 1977, \aap, 54, 661 

\bibitem[Tyson et al.(1984)]{tyson84}
Tyson, J. A., Valdes, F., Jarvis, J. F. \& Mills, A. P. 1984, \apjl, 281, L59

\bibitem[Tytler, Fan, \& Burles(1996)]{tytler96} 
Tytler, D., Fan, X.M., \& Burles, S. 1996, Nature, 381, 207

\bibitem[Weinberg \& Cole(1992)]{weinberg92}
Weinberg, D. H., \& Cole, S. 1992, \mnras, 259, 652

\bibitem[Weinberg et al.(1997)]{weinberg97}
Weinberg, D. H., Hernquist, L., \& Katz, N. 1997, \apj, 477, 8

\bibitem[Weinberg et al.(2002)]{weinberg02} 
Weinberg, D. H., Hernquist, L., \& Katz, N. 2002, \apj, 571, 15

\bibitem[Weinberg et al.(1998b)]{weinberg98b}
Weinberg, D. H., Katz, N., \& Hernquist, L. 1998b,
in Origins, eds. J. M. Shull, C. E. Woodward, \& H. Thronson,
(ASP: San Francisco), 21

\bibitem[Weinberg et al.(1999)]{weinberg99}
Weinberg, D. H., Dav\'e, R., Gardner,
J. P., Hernquist, L., \& Katz, N. 1999 in Photometric Redshifts and High
Redshift Galaxies, eds. R.  Weymann, L. Storrie-Lombardi, M. Sawicki \&
R. Brunner (ASP: San Francisco), 341341341

\bibitem[White, Hernquist, \& Springel(2001)]{white01}
White, M., Hernquist, L., \& Springel, V. 2001, unpublished, astro-ph/0107023

\bibitem[White et al.(1987)]{white87} 
White, S. D. M., Davis, M., 
Efstathiou, G., \& Frenk, C. S. 1987, Nature, 330, 451

\bibitem[White, Frenk, \& Davis(1983)]{white83}
White, S. D. M., Frenk, C. S., \& Davis, M. 1983, \apj, 274, L1


\bibitem[Willmer, da Costa \& Pellegrini(1998)]{willmer98}
Willmer, C. N. A., da Costa, L. N., \& Pellegrini, P. S. 1998, \aj, 115, 869

\bibitem[Wilson, Kaiser, \& Luppino(2001a)]{wilson01a}
Wilson, G., Kaiser, N., \& Luppino, G.~A.\ 2001a, \apj, 556, 601

\bibitem[Wilson et al.(2001b)]{wilson01b}
Wilson, G., Kaiser, N., Luppino, G. A., \& Cowie, L. L. 2001b,
\apj, 555, 572

\bibitem[Yang et al.(2002)]{yang02}
Yang, X., Mo, H. J., Kauffmann, G., \& Chu, Y.\ 2002, \mnras, submitted,
astro-ph/0205546

\bibitem[Yoshikawa et al.(2001)]{yoshikawa01} 
Yoshikawa, K., Taruya, A., Jing, Y. P., \& Suto, Y. 2001, \apj, 558, 520

\bibitem[Zehavi et al.(2002)]{zehavi02} 
Zehavi, I., et al. 2002, \apj, 571, 172


\end{thebibliography}
\end{document}